\PassOptionsToPackage{unicode}{hyperref}
\PassOptionsToPackage{hyphens}{url}
\documentclass[
  10pt,
]{article}
\usepackage{amsmath,amssymb}
\usepackage{iftex}
\ifPDFTeX
  \usepackage[T1]{fontenc}
  \usepackage[utf8]{inputenc}
  \usepackage{textcomp} 
\else 
  \usepackage{unicode-math} 
  \defaultfontfeatures{Scale=MatchLowercase}
  \defaultfontfeatures[\rmfamily]{Ligatures=TeX,Scale=1}
\fi
\usepackage{lmodern}
\ifPDFTeX\else
\fi
\IfFileExists{upquote.sty}{\usepackage{upquote}}{}
\IfFileExists{microtype.sty}{
  \usepackage[]{microtype}
  \UseMicrotypeSet[protrusion]{basicmath} 
}{}
\makeatletter
\@ifundefined{KOMAClassName}{
  \IfFileExists{parskip.sty}{%
    \usepackage{parskip}
  }{
    \setlength{\parindent}{0pt}
    \setlength{\parskip}{6pt plus 2pt minus 1pt}}
}{
  \KOMAoptions{parskip=half}}
\makeatother
\usepackage{xcolor}
\usepackage[margin=1in]{geometry}
\usepackage{longtable,booktabs,array}
\usepackage{calc} 
\usepackage{etoolbox}
\makeatletter
\patchcmd\longtable{\par}{\if@noskipsec\mbox{}\fi\par}{}{}
\makeatother
\IfFileExists{footnotehyper.sty}{\usepackage{footnotehyper}}{\usepackage{footnote}}
\makesavenoteenv{longtable}
\usepackage{graphicx}
\makeatletter
\def\maxwidth{\ifdim\Gin@nat@width>\linewidth\linewidth\else\Gin@nat@width\fi}
\def\maxheight{\ifdim\Gin@nat@height>\textheight\textheight\else\Gin@nat@height\fi}
\makeatother
\setkeys{Gin}{width=\maxwidth,height=\maxheight,keepaspectratio}
\makeatletter
\def\fps@figure{htbp}
\makeatother
\providecommand{\tightlist}{%
  \setlength{\itemsep}{0pt}\setlength{\parskip}{0pt}}
\ifLuaTeX
  \usepackage{selnolig}  
\fi
\IfFileExists{bookmark.sty}{\usepackage{bookmark}}{\usepackage{hyperref}}
\IfFileExists{xurl.sty}{\usepackage{xurl}}{} 
\hypersetup{
  hidelinks,
  pdfcreator={LaTeX via pandoc}}

\title{An Anchored Logistic Family for Bounded Trait Measurement and Growth}
\usepackage{etoolbox}
\makeatletter
\providecommand{\subtitle}[1]{
  \apptocmd{\@title}{\par {\large #1 \par}}{}{}
}
\makeatother
\subtitle{Origin Before Unit}
\author{Jaehwa Choi\thanks{Department of Educational Leadership, Graduate School of Education and Human Development, The George Washington University, Washington, DC 20010, USA. Email: \texttt{jaechoi@gwu.edu}. ORCID: \url{https://orcid.org/0000-0003-4225-9582}.}}
\date{August 2026}

\begin{document}
\maketitle

\hypertarget{abstract}{%
\subsection{Abstract}\label{abstract}}

Latent trait models differ less in what they measure than in what they
fix. Item response theory frees the trait from the items administered
but, in doing so, surrenders the origin and unit of the scale to
convention. The Cognitive Trait Model (CTM; Choi, 2022) restores both by
bounding the trait on {[}0, 1{]}, where 0 denotes an ignorance level and
1 a mastery level defined by the task domain. We generalize the
derivation behind CTM into a family indexed by the mastery anchor
\emph{L}: the published model is the case \emph{L} = 1, while the limit
\emph{L} \(\to\) \(\infty\) gives a half-truncated link with an absolute
origin but no ceiling. \textbf{Origin before unit.}

Bounded support admits Gauss--Legendre quadrature without truncation
error, so item and person parameters follow from Bayes modal EM rather
than MCMC. On identical data and priors this reproduces the published
estimates (\(\theta\) correlation 0.9999) while running 35 times faster
than a converged random-walk sampler and 73 times faster than NUTS; a
single adaptive-testing update costs about five microseconds. Applied to
the time axis, the same link becomes a four-parameter growth curve
spanning accelerating, decelerating, plateauing and declining
trajectories, though not non-monotone ones. The family reduces to a
reparameterization of the logistic model only when all items share a
slope, so a measurement advantage is possible in principle; three
simulations find none, locating the contribution in interpretation and
computation rather than in measurement itself. In the LSAT reanalysis,
examinees with perfect scores read 67\%, 75\% or 100\% of mastery
depending on the estimator --- the criterion-referenced claim is
meaningful only with its uncertainty attached, and the coordinate is not
in general the proportion of the domain a person can perform.

\textbf{Keywords:} cognitive trait model; item response theory; bounded
latent variable; Gauss--Legendre quadrature; Bayes modal estimation;
adaptive testing; learning trajectory; criterion-referenced measurement

\begin{center}\rule{0.5\linewidth}{0.5pt}\end{center}

\hypertarget{introduction}{%
\section{1. Introduction}\label{introduction}}

\begin{center}\rule{0.5\linewidth}{0.5pt}\end{center}

\hypertarget{what-a-measurement-model-fixes}{%
\subsection{1.1 What a measurement model
fixes}\label{what-a-measurement-model-fixes}}

The relation between what a person does on a set of tasks and what we
take that to reveal about them has been studied for over a century ---
since Spearman (1904) --- and the arc of that study can be told as a
sequence of decisions about what to hold still.

The starting point was as simple as it could be: \textbf{some answers
were right and some were wrong.} Counting them gave a score with two
natural endpoints, zero and the number of items, and classical test
theory built a serviceable and durable apparatus on that foundation
(Lord \& Novick, 1968). The score's weakness is equally simple. Its
units are the items administered. Eighty percent on an easy form and
eighty percent on a hard one are different quantities wearing the same
number, and no amount of care in computing reliability repairs that.

Item response theory removed the dependence by modelling the items
explicitly. Difficulty and discrimination became parameters rather than
features of the scale, and a person's ability could be spoken of
independently of the particular form they took. This is among the most
consequential ideas in psychometrics, and we take it as settled.

But a scale freed from the items is not thereby anchored to anything
else. The likelihood is unchanged if the trait is subjected to any
strictly increasing transformation and the item parameters are
transformed to match, so an origin and a unit have to be supplied from
outside the data. The convention that supplies them --- mean zero and
standard deviation one in the calibration sample --- is a statement
about \textbf{the other people in the sample}. A trait level of
\(\theta\) = 0 identifies someone at the average of a reference group;
it says nothing about what they can do.

For summative and comparative purposes that is entirely adequate;
ranking is what is wanted. For formative assessment (Black \& Wiliam,
1998), for learning progressions, for adaptive instruction --- settings
in which the question is not \emph{who is ahead} but \emph{how far along
is this learner, and how far do they have to go} --- it is not. The
practitioner's question is criterion-referenced and the metric is
norm-referenced, and the gap between them is bridged, when it is bridged
at all, by transformations applied after the fact.

Choi (2022) proposed closing the gap at the source by asking what the
numbers should mean before asking how to estimate them. If the domain of
tasks is specified, then there is a level at which none of them would be
performed successfully and a level at which all of them would be. Call
these the \emph{ignorance} and \emph{mastery} levels, assign them 0 and
1, and the trait is bounded on the unit interval with a reading that
requires no further translation: \textbf{\(\theta\) = 0.62 is 62\% of
mastery}, a reading examined in \S{}4.5. The Cognitive Trait Model gives
the link function that makes this well defined, derives it from the
two-parameter logistic function, and demonstrates estimation by Markov
chain Monte Carlo. The idea of a bounded latent proportion has
antecedents. The beta-binomial models of criterion-referenced testing
place a true score directly on {[}0, 1{]} (Wilcox, 1981); a
continuous-response line within item response theory bounds the response
instead, from Samejima's (1973) continuous response model to the beta
response model of Noel and Dauvier (2007); and diagnostic classification
models supply a discrete counterpart, assigning mastery as a status
rather than a level (von Davier \& Lee, 2019). The Cognitive Trait Model
differs from the first in modelling the items, from the second in
bounding the trait rather than the response, and from the third in
keeping the continuum.

That the question of an origin should arrive last in this sequence has a
precedent. Counting numbers are as old as record-keeping; zero is not.
It appeared first as a placeholder in Babylonian positional notation and
became a number in its own right only in seventh-century India,
millennia after the numbers it anchors (Kaplan, 2000; Ifrah, 2000) ---
and when it did arrive, it paid twice. It gave the number system a
meaning for \emph{none}, and it made positional arithmetic, and with it
efficient computation, possible. An origin comes late because nothing in
the practice of counting forces it; it is valuable because, once fixed,
both the interpretation and the computation are built on it. The
argument of this paper is that the same double payment occurs here.

Three transitions, then, and none of them adds information to the data.
Each relocates a dependence: classical test theory depends on the items,
item response theory on the population, the Cognitive Trait Model on the
task domain. This observation is the frame within which our own results
should be read, and we return to it in \S{}4.

\begin{center}\rule{0.5\linewidth}{0.5pt}\end{center}

\hypertarget{dependence-does-not-vanish-it-can-be-declared}{%
\subsection{1.2 Dependence does not vanish; it can be
declared}\label{dependence-does-not-vanish-it-can-be-declared}}

Since no test-based measurement escapes dependence, the useful question
is not which model is free of it but \textbf{where each one puts it, and
whether the placement is visible.}

\begin{longtable}[]{@{}
  >{\raggedright\arraybackslash}p{(\columnwidth - 4\tabcolsep) * \real{0.3333}}
  >{\raggedright\arraybackslash}p{(\columnwidth - 4\tabcolsep) * \real{0.3333}}
  >{\raggedright\arraybackslash}p{(\columnwidth - 4\tabcolsep) * \real{0.3333}}@{}}
\toprule\noalign{}
\begin{minipage}[b]{\linewidth}\raggedright
\end{minipage} & \begin{minipage}[b]{\linewidth}\raggedright
dependence resides in
\end{minipage} & \begin{minipage}[b]{\linewidth}\raggedright
is it stated?
\end{minipage} \\
\midrule\noalign{}
\endhead
\bottomrule\noalign{}
\endlastfoot
Classical test theory & the items administered & no; the score is
reported as a number \\
Item response theory & the calibration population & no; hidden by the
convention \(\mu\) = 0, \(\sigma\) = 1 \\
Cognitive Trait Model & the task domain & \textbf{yes; it is in the
definition} \\
\end{longtable}

Item response theory appears to escape the item dependence of classical
test theory, and in one sense it does. What it actually does is move the
dependence to the population, and that move is not recorded anywhere in
the reported score. A \(\theta\) of 1.2 does not announce which sample
it is 1.2 standard deviations above.

The bounded model does not remove the dependence either. Its \(\theta\)
= 1 means ``all tasks in \emph{this} domain'', and a different domain is
a different scale. The difference is that it says so. A declared
dependence can be managed, audited, and reproduced; ``62\% of mastery of
the following twenty competencies'' is a claim that can be checked and,
if wrong, shown to be wrong. \textbf{This explicitness, rather than any
improvement in precision, is what the bounded scale offers}, and Section
4 of this paper is largely an attempt to establish that claim by failing
to find anything else.

We should say at once that \S{}4.5 finds the Cognitive Trait Model
carrying an undeclared dependence of its own, on the calibration
population, exactly as item response theory does. The remedy is not to
abandon the framework but to carry its own logic through: to anchor
explicitly what the construction only names.

\begin{center}\rule{0.5\linewidth}{0.5pt}\end{center}

\hypertarget{from-interpretation-to-use}{%
\subsection{1.3 From interpretation to
use}\label{from-interpretation-to-use}}

The interpretive case for a bounded trait was made in the source article
and we do not repeat it. Our question is what stands between that case
and its use in the settings it was designed for.

Adaptive instruction imposes two requirements that a post hoc analysis
does not.

The first is \textbf{speed}. A system that selects the next task in
response to the last one must re-estimate between items. Markov chain
Monte Carlo, which the source article uses, is not merely inconvenient
at that timescale --- it is the wrong kind of algorithm for it,
requiring thousands of draws and convergence diagnostics to answer a
question that recurs every few seconds.

The second is \textbf{honest uncertainty}. Deciding when a learner has
demonstrated mastery, when to stop testing, and which task to present
next are all decisions that consume the standard error rather than the
point estimate. A model that reports a precision it does not have will
stop too early and select badly, and the errors will not announce
themselves.

These requirements are what this paper addresses. The central
observation is that the property responsible for the interpretive claim
--- the bounded support --- is also what makes both requirements
satisfiable. Because \(\theta\) lives on a closed interval, quadrature
is exact rather than truncated, the item and person parameters follow
from a deterministic algorithm rather than a sampler, and the grid on
which everything is computed is fixed once for the lifetime of an item
bank. The interpretation and the computation are consequences of the
same decision --- the anchoring, like zero, pays twice.

\begin{center}\rule{0.5\linewidth}{0.5pt}\end{center}

\hypertarget{contributions}{%
\subsection{1.4 Contributions}\label{contributions}}

\textbf{A one-parameter family.} The derivation underlying the Cognitive
Trait Model imposes two boundary conditions, one fixing the origin and
one fixing the unit. Imposing only the first yields a family indexed by
the mastery anchor \emph{L}: the published model is \emph{L} = 1, and
\emph{L} \(\to\) \(\infty\) gives a link with an absolute origin and no
ceiling. \textbf{Origin before unit}, since a criterion-referenced scale
cannot do without an origin whereas the unit is a further and stronger
commitment (\S{}2).

\textbf{Estimation without MCMC.} Bayes modal estimation on a
Gauss--Legendre grid reproduces the published estimates to a correlation
of 0.9999 while running 35 times faster than a converged random-walk
sampler and 73 times faster than the No-U-Turn Sampler, and reduces a
single adaptive-testing update to roughly five microseconds (\S{}3).

\textbf{The information function}, listed as future work in the source
article, together with a characterization of its divergence at the
mastery boundary and the conditions under which that divergence does and
does not matter in practice (\S{}3.5).

\textbf{An empirical comparison with item response theory}, which we
report as null: three separate attempts to find a measurement advantage
found none. We also show that the family is a reparameterization of the
logistic model only when all items share a common slope, so the null is
a finding rather than a tautology (\S{}4).

\textbf{A sequential extension.} The link function, applied to the time
axis, is a four-parameter growth curve covering accelerating,
decelerating, plateauing and declining trajectories in a single form.
Fitted to a learner's own history it supplies the prior for the next
occasion. We document where this fails --- naive carrying-forward is
worse than no memory at all, and non-monotone trajectories cannot be
represented --- and the discipline required to make it work (\S{}5).

\textbf{Three corrections} to the source article, concerning a limit
stated in the text, an inference about parameter stability, and a
misaligned table (\S{}2.7).

\begin{center}\rule{0.5\linewidth}{0.5pt}\end{center}

\hypertarget{scope-and-what-we-do-not-claim}{%
\subsection{1.5 Scope and what we do not
claim}\label{scope-and-what-we-do-not-claim}}

We compare against item response theory throughout, and we should be
explicit about what the comparison is and is not.

The speed results are \textbf{against the estimation path the source
article uses}. Item response theory has had marginal maximum likelihood
for four decades (Bock \& Aitkin, 1981) and is not slow; nothing here is
a claim to the contrary.

The measurement comparison is \textbf{null and we report it as such}.
Where differences appear (\S{}4.4) they are small, condition-dependent,
and run in both directions. We do not claim that the bounded model
measures better.

The growth results assume \textbf{monotone trajectories and anchored
item parameters}. Both assumptions are examined rather than asserted,
and the failures under violation are reported (\S{}5.5).

We use the unbounded member of the family as a growth curve only, not as
a measurement model. Doing both would put two competing readings of the
trait in one paper, and the choice between them is theoretical rather
than empirical.

\begin{center}\rule{0.5\linewidth}{0.5pt}\end{center}

\hypertarget{the-anchored-logistic-family}{%
\section{2. The Anchored Logistic
Family}\label{the-anchored-logistic-family}}

\begin{center}\rule{0.5\linewidth}{0.5pt}\end{center}

\hypertarget{what-a-scale-must-fix}{%
\subsection{2.1 What a scale must fix}\label{what-a-scale-must-fix}}

A latent trait scale is not determined by the data alone. For any
strictly increasing map \emph{g}, the pair (\emph{g}(\(\theta\)), items
reparameterized through \emph{g}) implies the same response
probabilities as (\(\theta\), items), so the likelihood cannot
distinguish them. Two degrees of freedom --- an origin and a unit ---
must be fixed by something outside the data.

Traditions differ in what they use for that purpose, and the choice, not
the mathematics, is what separates the models.

\begin{longtable}[]{@{}
  >{\raggedright\arraybackslash}p{(\columnwidth - 6\tabcolsep) * \real{0.2500}}
  >{\raggedright\arraybackslash}p{(\columnwidth - 6\tabcolsep) * \real{0.2500}}
  >{\raggedright\arraybackslash}p{(\columnwidth - 6\tabcolsep) * \real{0.2500}}
  >{\raggedright\arraybackslash}p{(\columnwidth - 6\tabcolsep) * \real{0.2500}}@{}}
\toprule\noalign{}
\begin{minipage}[b]{\linewidth}\raggedright
\end{minipage} & \begin{minipage}[b]{\linewidth}\raggedright
origin
\end{minipage} & \begin{minipage}[b]{\linewidth}\raggedright
unit
\end{minipage} & \begin{minipage}[b]{\linewidth}\raggedright
fixed by
\end{minipage} \\
\midrule\noalign{}
\endhead
\bottomrule\noalign{}
\endlastfoot
Classical test theory & 0 correct & 1 item & the items administered \\
Item response theory & sample mean & sample SD & convention \\
\textbf{CTM} (\emph{L} = 1) & ignorance level & mastery level & the task
domain \\
\textbf{HCTM} (\emph{L} \(\to\) \(\infty\)) & ignorance level & ---
(none) & the task domain \\
\end{longtable}

Classical test theory ties both to the particular items given: 80\%
correct means one thing on an easy form and another on a hard one. Item
response theory severs that tie by modelling item properties explicitly,
but the freed scale must still be pinned somewhere, and the customary
choice --- mean 0 and standard deviation 1 in the calibration sample ---
pins it to the \emph{population} rather than to the construct. A trait
level of \(\theta\) = 0 is a statement about other examinees.

Choi (2022) proposed pinning both to the construct instead. If the
domain of tasks is specified, there exists a level at which none of them
is expected to be performed successfully (an \emph{ignorance level}) and
a level at which all of them are (a \emph{mastery level}). Assigning 0
and 1 to these levels yields a trait bounded on {[}0, 1{]} that reads as
a proportion of mastery.

What is being claimed when a level is called an origin deserves care,
because the two traditions do not disagree about where to put a number;
they disagree about what kind of thing the zero of a trait scale is. The
conventional \(\theta\) = 0 is \emph{indexical}: it points at whoever
happened to be calibrated --- the average member of this sample --- and
points at someone else the moment the sample changes. The ignorance
level is \emph{definitional}: it refers to the task domain, and it means
the same thing whoever is measured, or whether anyone is measured at
all. Only the second kind of zero can support a criterion-referenced
statement.

The definition has three parts, and the existence part is the least of
them. A level below which no task in the domain succeeds is implied by
any monotone response model --- including the logistic one, where it
sits at \(\theta\) = \(- \infty\). The floor of item response theory is
an asymptote: approached, never occupied, and therefore useless as an
origin, since no distance can be measured from a point that cannot be
reached. What the bounded construction contributes is not the existence
of an ignorance level but its \emph{relocation} --- from an asymptote to
a finite, attainable point of the scale. That is the literal content of
the word \emph{truncated} in the derivation that follows.

The second part is why a single point, and here the answer is
identifiability rather than assumption. Below the ignorance level,
examinees are observationally exchangeable: every task fails, the data
are identical, and --- the same principle that governs the ceiling in
\S{}4.2 --- no model can separate examinees whose data are identical. A
scale should not distinguish what no observation could; collapsing the
region to one point asserts nothing about the people in it. The origin
is then the \emph{edge} of that region: the lowest level at which a
success begins to carry information, and Proposition 2.3 will show that
the model honours the definition --- the derivative of the response
function at 0 is finite and strictly positive, so information begins at
the origin rather than accumulating asymptotically toward it.

The third part is the naming: assigning the number 0 to this level is a
definition, not a discovery, and the three-parameter member shows what
the definition is actually about. With guessing, observed success at the
ignorance level is not zero but \(\gamma\); what vanishes at the origin
is \emph{trait-caused} success, and \(\gamma\) is the model's own
estimate of the success owed to the response format rather than to the
trait. The origin, precisely stated, is the level at which the trait
contributes nothing to performance. How that level is \emph{found} ---
as opposed to defined --- is an operational question that the constraint
P(0) = 0 answers only at calibration, and \S{}4.5 will show that the
answer is not yet sufficient: under marginal estimation it is the prior,
not the constraint, that ends up holding the scale. The distance between
the definition and its operationalization is one of the quantities this
paper measures.

We take the proposal seriously enough to ask what follows when only
\emph{one} of the two anchors is imposed. \textbf{Origin before unit}:
the origin is the anchor a criterion-referenced scale cannot do without,
whereas the unit is an additional and stronger commitment.

\begin{center}\rule{0.5\linewidth}{0.5pt}\end{center}

\hypertarget{the-published-derivation}{%
\subsection{2.2 The published
derivation}\label{the-published-derivation}}

Choi (2022) begins with the two-parameter logistic function (Birnbaum,
1968)

\[z(\theta) = \sigma\!\big(\alpha(\theta-\beta)\big), \qquad
\sigma(u) = \frac{e^{u}}{1+e^{u}},\]

shifts it down by \(\tfrac12\), rescales by \(\delta\), and translates
by \(\varepsilon\):

\[P(\theta) \;=\; \delta\!\left[\sigma\!\big(\alpha(\theta-\beta)\big) - \tfrac12\right] + \varepsilon .
\tag{2.1}\]

Requiring \emph{P}(0) = 0 and \emph{P}(1) = 1 determines (\(\delta\),
\(\varepsilon\)) and yields, after simplification, his Eq. (13):

\[P_{2}(\theta) \;=\;
\frac{1-e^{\alpha\theta}}{1+e^{\alpha(\theta-\beta)}}\cdot
\frac{1+e^{\alpha(1-\beta)}}{1-e^{\alpha}} .
\tag{2.2}\]

A lower asymptote \(\gamma\) is then reintroduced as in the 3PL model
(Birnbaum, 1968), \emph{P}\(_{3}\) = \(\gamma\) + (1 \(-\)
\(\gamma\))\emph{P}\(_{2}\) (his Eq. 12).

Two observations organize what follows. First, (1 \(-\)
\emph{e}\(^{\alpha \theta }\)) and (1 \(-\) \emph{e}\(^{\alpha }\)) are
both negative for \(\alpha\) \textgreater{} 0, so their ratio is
positive and equals expm1(\(\alpha\)\(\theta\))/expm1(\(\alpha\));
writing (2.2) this way avoids a sign error that is easy to make in
implementation. Second, and this is the point of departure, the two
boundary conditions are not equally necessary. \emph{P}(0) = 0 fixes the
origin; \emph{P}(1) = 1 fixes the unit. The upper condition can be
relaxed without destroying the construction.

\begin{center}\rule{0.5\linewidth}{0.5pt}\end{center}

\hypertarget{generalization-an-arbitrary-mastery-anchor}{%
\subsection{2.3 Generalization: an arbitrary mastery
anchor}\label{generalization-an-arbitrary-mastery-anchor}}

Let \emph{L} \textgreater{} 0 denote the trait level at which every task
in the domain is expected to be performed successfully, and impose
\emph{P}(0) = 0 and \emph{P}(\emph{L}) = 1 on (2.1). Subtracting the two
resulting equations eliminates \(\varepsilon\) and gives

\[\delta\left[\sigma\!\big(\alpha(L-\beta)\big) - \sigma(-\alpha\beta)\right] = 1 ,\]

so that \(\delta\) is determined and (2.1) collapses to a
\textbf{normalized logistic}:

\[\boxed{\;P_{L}(\theta) \;=\;
\frac{\sigma\!\big(\alpha(\theta-\beta)\big) - \sigma(-\alpha\beta)}
{\sigma\!\big(\alpha(L-\beta)\big) - \sigma(-\alpha\beta)}\;}
\tag{2.3}\]

Equation (2.3) is the form we work with throughout: it makes every
property in \S{}2.5 immediate, and it is numerically better behaved than
the algebraically equivalent expansion

\[P_{L}(\theta) \;=\;
\frac{e^{\alpha\theta}-1}{1+e^{\alpha(\theta-\beta)}}
\cdot
\frac{1+e^{\alpha(L-\beta)}}{e^{\alpha L}-1} .
\tag{2.4}\]

The kernel of (2.4) is free of \emph{L}; only the normalizer depends on
it. The family is thus indexed by a single quantity with a substantive
meaning --- \emph{where the mark ``1'' is placed on the ruler} --- and
its members differ by a multiplicative constant, not by shape.

\textbf{Member 1 (\emph{L} = 1).} Substituting \emph{L} = 1 returns Eq.
(2.2) exactly; the published model is a member of the family rather than
a separate object. \emph{(Maximum absolute difference between (2.3) with
L = 1 and the transcribed Eq. (13), over \(\alpha\) \(\in\) \{0.5,
\ldots, 50\} \(\times\) \(\beta\) \(\in\) \{0.05, \ldots, 0.95\} on a
401-point \(\theta\) grid: 7.3 \(\times\) 10\(^{-}\)\(^{1}\)\(^{5}\).)}

\textbf{Member 2 (\emph{L} \(\to\) \(\infty\)).} Since
\(\sigma\)(\(\alpha\)(\emph{L} \(-\) \(\beta\))) \(\to\) 1 and 1 \(-\)
\(\sigma\)(\(- \alpha\)\(\beta\)) = \(\sigma\)(\(\alpha\)\(\beta\)),

\[\boxed{\;P_{\infty}(\theta) \;=\;
\frac{\sigma\!\big(\alpha(\theta-\beta)\big) - \sigma(-\alpha\beta)}{\sigma(\alpha\beta)}
\;=\;
\frac{e^{\alpha\theta}-1}{e^{\alpha\beta}\left(1+e^{\alpha(\theta-\beta)}\right)}\;}
\tag{2.5}\]

We call (2.5) the \textbf{half-truncated CTM (HCTM)}: truncated at the
origin, unbounded above. It follows directly from imposing the single
condition \emph{P}(0) = 0 and letting the upper asymptote be the one the
logistic already has.

Convergence of (2.4) to (2.5) is governed by \(\alpha\)(\emph{L} \(-\)
\(\beta\)) rather than by \emph{L} alone: for \(\alpha\) = 1 the maximum
absolute difference over \(\theta\) \(\in\) {[}0.01, 0.99{]} is 2
\(\times\) 10\(^{-}\)\(^{9}\) at \emph{L} = 20 and 7 \(\times\)
10\(^{-}\)\(^{1}\)\(^{6}\) at \emph{L} = 50, whereas for \(\alpha\) = 5
it is already 6 \(\times\) 10\(^{-}\)\(^{1}\)\(^{6}\) at \emph{L} = 10.
The expanded form (2.4) should not be used for large \emph{L}, since
\emph{e}\(^{\alpha L}\) overflows double precision once
\(\alpha\)\emph{L} \textgreater{} 709; the normalized form (2.3) and the
closed form (2.5) have no such restriction.

Figure 1 displays the family on a common kernel (\(\alpha\) = 1.2,
\(\beta\) = 0.5): the members share one curve and differ only in where
the mark `1' falls, and the \emph{L} = 5 member is already visually
indistinguishable from the \emph{L} \(\to\) \(\infty\) limit.

\includegraphics[width=1\textwidth,height=\textheight]{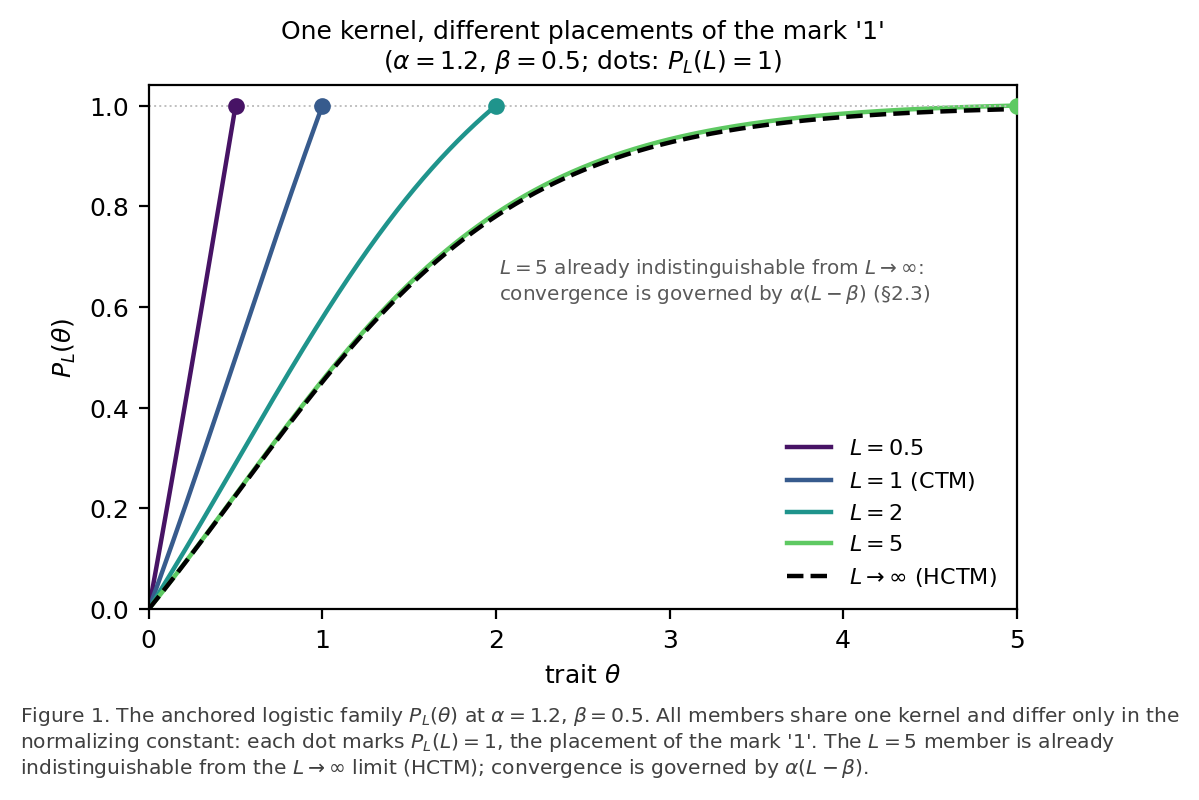}

\begin{center}\rule{0.5\linewidth}{0.5pt}\end{center}

\hypertarget{what-each-member-claims}{%
\subsection{2.4 What each member claims}\label{what-each-member-claims}}

\begin{longtable}[]{@{}
  >{\raggedright\arraybackslash}p{(\columnwidth - 8\tabcolsep) * \real{0.2000}}
  >{\raggedright\arraybackslash}p{(\columnwidth - 8\tabcolsep) * \real{0.2000}}
  >{\raggedright\arraybackslash}p{(\columnwidth - 8\tabcolsep) * \real{0.2000}}
  >{\raggedright\arraybackslash}p{(\columnwidth - 8\tabcolsep) * \real{0.2000}}
  >{\raggedright\arraybackslash}p{(\columnwidth - 8\tabcolsep) * \real{0.2000}}@{}}
\toprule\noalign{}
\begin{minipage}[b]{\linewidth}\raggedright
\end{minipage} & \begin{minipage}[b]{\linewidth}\raggedright
support
\end{minipage} & \begin{minipage}[b]{\linewidth}\raggedright
fixed
\end{minipage} & \begin{minipage}[b]{\linewidth}\raggedright
scale type
\end{minipage} & \begin{minipage}[b]{\linewidth}\raggedright
\(\theta\) = \emph{L} means
\end{minipage} \\
\midrule\noalign{}
\endhead
\bottomrule\noalign{}
\endlastfoot
CTM (\emph{L} = 1) & {[}0, 1{]} & origin + unit & absolute & full
mastery of the domain \\
HCTM (\emph{L} \(\to\) \(\infty\)) & {[}0, \(\infty\)) & origin only &
ratio & (nothing; no ceiling) \\
\end{longtable}

HCTM retains a true zero --- the ignorance level is an attainable point
of the scale rather than an asymptote --- while declining to assert that
any finite level exhausts the domain. Ratios are therefore meaningful,
but proportions of mastery are not.

The slogan of \S{}2.1 --- origin before unit --- can now be given its
grounds, and they are three. The first is the hierarchy of scale types
visible in the table. An origin alone already licenses ratio statements
(``twice as far from ignorance''); adding a unit upgrades them to
absolute ones (``62\% of the way from ignorance to mastery''). The
minimal requirement of a criterion-referenced reading is the former, and
the latter is a refinement of it, not a precondition for it. Whether
that coordinate is also the proportion of the domain a person can
perform is a separate question, taken up in \S{}4.5. The second is the
asymmetry of the commitments themselves. The unit anchor is a
completeness claim: that the tasks in hand exhaust the domain, so that
some finite level performs \emph{all} of it --- and the source article
itself observes that uncertainty at the top of the domain propagates
through the entire metric. The origin makes no such claim; an ignorance
level requires only that every task can be failed, which no future
enlargement of the domain threatens. (Whether either anchor survives
\emph{estimation} is a separate, operational question; \S{}4.5 answers
it, and the exposure it finds runs through the population, not the
domain.) The third is the time axis. A growth curve needs a floor to
grow from and should not be assigned a ceiling in advance (\S{}5.3);
that a member with an origin and no unit exists at all --- and is
precisely the member Section 5 uses --- is the constructive form of the
priority claim. This is also the asymmetry behind the aside of \S{}1.1:
the anchor that arrives last is the one nothing in the data forces, and
the one on which the new readings depend.

We use \emph{L} = 1 as the \textbf{measurement} model, where the
proportion-of-mastery reading is the object of interest, and \emph{L}
\(\to\) \(\infty\) as the \textbf{growth} model on the time axis
(Section 5), where no finite horizon should be imposed. We do not fit
HCTM as a measurement model: the two readings would compete, and the
choice between them is theoretical rather than empirical.

\begin{center}\rule{0.5\linewidth}{0.5pt}\end{center}

\hypertarget{properties}{%
\subsection{2.5 Properties}\label{properties}}

Throughout \(\alpha\) \textgreater{} 0 and 0 \textless{} \(\beta\)
\textless{} \emph{L}. Write \emph{s}(\(\theta\)) =
\(\sigma\)(\(\alpha\)(\(\theta\) \(-\) \(\beta\))) and \emph{D} =
\(\sigma\)(\(\alpha\)(\emph{L} \(-\) \(\beta\))) \(-\)
\(\sigma\)(\(- \alpha\)\(\beta\)), so that (2.3) reads \emph{P\(_{L}\)}
= {[}\emph{s}(\(\theta\)) \(-\)
\(\sigma\)(\(- \alpha\)\(\beta\)){]}/\emph{D}.

\textbf{Proposition 2.1 (boundaries).} \emph{P\(_{L}\)}(0) = 0 and
\emph{P\(_{L}\)}(\emph{L}) = 1; for the three-parameter version
\emph{P}\(_{3}\)(0) = \(\gamma\) and \emph{P}\(_{3}\)(\emph{L}) = 1.

\emph{Proof.} At \(\theta\) = 0, \emph{s}(0) =
\(\sigma\)(\(- \alpha\)\(\beta\)), so the numerator of (2.3) vanishes.
At \(\theta\) = \emph{L}, \emph{s}(\emph{L}) =
\(\sigma\)(\(\alpha\)(\emph{L} \(-\) \(\beta\))), so numerator and
denominator coincide. The three-parameter statements follow from
\emph{P}\(_{3}\) = \(\gamma\) + (1 \(-\) \(\gamma\))\emph{P}\(_{2}\).
\(\square\)

\textbf{Proposition 2.2 (strict monotonicity).} \emph{P\(_{L}\)} is
strictly increasing on \(\mathbb{R}\), and \emph{D} \textgreater{} 0.

\emph{Proof.} Since \(\alpha\) \textgreater{} 0 and \emph{L}
\textgreater{} \(\beta\), we have \(\alpha\)(\emph{L} \(-\) \(\beta\))
\textgreater{} 0 \textgreater{} \(- \alpha\)\(\beta\); \(\sigma\) is
strictly increasing, hence \emph{D} \textgreater{} 0. Then (2.3) is an
increasing affine transform of the strictly increasing function
\emph{s}, so \emph{P\(_{L}\)} is strictly increasing. \(\square\)

\emph{(Numerical check: the minimum first difference over a 2001-point
grid is \(- 4.4\) \(\times\) 10\(^{-}\)\(^{1}\)\(^{6}\), one unit in the
last place at saturation.)}

\textbf{Proposition 2.3 (derivative).} For all \(\theta\) \(\in\)
\(\mathbb{R}\),

\[P_{L}'(\theta) \;=\; \frac{\alpha}{D}\,s(\theta)\big[1-s(\theta)\big],
\qquad P_{3}' = (1-\gamma)P_{L}' ,
\tag{2.6}\]

which is finite and strictly positive everywhere, including at
\(\theta\) = 0 and \(\theta\) = \emph{L}.

\emph{Proof.} Differentiate (2.3) and use \(\sigma\)\('\)(\emph{u}) =
\(\sigma\)(\emph{u}){[}1 \(-\) \(\sigma\)(\emph{u}){]} with the chain
rule. \(\square\)

\emph{Remark.} The expanded form (2.4) yields the algebraically
equivalent expression \emph{P}\('\) =
\emph{P}\(\,\cdot\,\)\(\alpha\)\(\,\cdot\,\){[}(1 \(-\)
\emph{e}\(^{-\alpha \theta }\))\(^{-}\)\(^{1}\) \(-\)
\emph{s}(\(\theta\)){]}, in which the bracket diverges as \(\theta\)
\(\to\) 0 while \emph{P} vanishes. That removable singularity is an
artefact of the expansion, not of the model; (2.6) has no special case.
\emph{(Both forms agree with central differences to \(\le\) 5 \(\times\)
10\(^{-}\)\(^{7}\), and with each other at \(\theta\) = 0 to eight
decimal places.)}

\textbf{Proposition 2.4 (limit as \(\alpha\) \(\to\) 0).}
\emph{P\(_{L}\)}(\(\theta\)) \(\to\) \(\theta\)/\emph{L} pointwise, and
hence \emph{P}\(_{3}\) \(\to\) \(\gamma\) + (1 \(-\)
\(\gamma\))\(\theta\)/\emph{L}. For \emph{L} = 1 this is \(\gamma\) + (1
\(-\) \(\gamma\))\(\theta\).

\emph{Proof.} Expand \(\sigma\)(\emph{u}) = \(\tfrac12\) + \emph{u}/4 +
\emph{O}(\emph{u}\(^{3}\)). The numerator of (2.3) is
\(\alpha\)(\(\theta\) \(-\) \(\beta\))/4 + \(\alpha\)\(\beta\)/4 +
\emph{O}(\(\alpha\)\(^{3}\)) = \(\alpha\)\(\theta\)/4 +
\emph{O}(\(\alpha\)\(^{3}\)) and the denominator is \(\alpha\)(\emph{L}
\(-\) \(\beta\))/4 + \(\alpha\)\(\beta\)/4 +
\emph{O}(\(\alpha\)\(^{3}\)) = \(\alpha\)\emph{L}/4 +
\emph{O}(\(\alpha\)\(^{3}\)). The ratio tends to \(\theta\)/\emph{L}.
\(\square\)

This differs from the expression given in the source text; see \S{}2.7.

\textbf{Proposition 2.5 (the location parameter is not a median, with an
exact exception).} \emph{P\(_{L}\)}(\(\beta\)) depends on \(\alpha\) in
general, but \emph{P\(_{L}\)}(\emph{L}/2) = \(\tfrac12\) exactly
whenever \(\beta\) = \emph{L}/2, for every \(\alpha\) and \emph{L}. For
HCTM the value at \(\beta\) is available in closed form,

\[P_{\infty}(\beta) = \tfrac12\left(1 - e^{-\alpha\beta}\right).\]

\emph{Proof.} At \(\theta\) = \(\beta\), \emph{s}(\(\beta\)) =
\(\tfrac12\), so \emph{P\(_{L}\)}(\(\beta\)) = {[}\(\tfrac12\) \(-\)
\(\sigma\)(\(- \alpha\)\(\beta\)){]}/\emph{D} =
{[}\(\sigma\)(\(\alpha\)\(\beta\)) \(-\) \(\tfrac12\){]}/\emph{D}. If
\(\beta\) = \emph{L}/2 then \(\sigma\)(\(\alpha\)(\emph{L} \(-\)
\(\beta\))) = \(\sigma\)(\(\alpha\)\(\beta\)) and
\(\sigma\)(\(- \alpha\)\(\beta\)) = 1 \(-\)
\(\sigma\)(\(\alpha\)\(\beta\)), so \emph{D} =
2\(\sigma\)(\(\alpha\)\(\beta\)) \(-\) 1 =
2{[}\(\sigma\)(\(\alpha\)\(\beta\)) \(-\) \(\tfrac12\){]}, giving
\emph{P} = \(\tfrac12\). For \emph{L} \(\to\) \(\infty\), \emph{D} =
\(\sigma\)(\(\alpha\)\(\beta\)) and \emph{P}\(_{\infty }\)(\(\beta\)) =
{[}\(\sigma\)(\(\alpha\)\(\beta\)) \(-\)
\(\tfrac12\){]}/\(\sigma\)(\(\alpha\)\(\beta\)) = 1 \(-\) (1 +
\emph{e}\(^{-\alpha \beta }\))/2. \(\square\)

\emph{(Verified: \textbar P(L/2) \(-\) \(\tfrac12\)\textbar{} \(\le\)
5.6 \(\times\) 10\(^{-}\)\(^{1}\)\(^{7}\) for \(\alpha\) \(\in\) \{0.5,
\ldots, 100\} and L \(\in\) \{1, 3\}; the HCTM identity to 2.8
\(\times\) 10\(^{-}\)\(^{1}\)\(^{6}\).)}

Numerically, for \emph{L} = 1:

\begin{longtable}[]{@{}llllll@{}}
\toprule\noalign{}
\(\beta \backslash \alpha\) & 1 & 5 & 10 & 25 & 100 \\
\midrule\noalign{}
\endhead
\bottomrule\noalign{}
\endlastfoot
0.1 & 0.106 & 0.200 & 0.316 & 0.459 & 0.500 \\
0.5 & 0.500 & 0.500 & 0.500 & 0.500 & 0.500 \\
0.9 & 0.894 & 0.800 & 0.684 & 0.541 & 0.500 \\
\end{longtable}

Figure 2 traces \emph{P}(\(\theta\) = \(\beta\)) against \(\alpha\)
across this range. Deviation from \(\tfrac12\) is largest when
\(\alpha\) is small and \(\beta\) is extreme, and vanishes as \(\alpha\)
grows. A practical consequence: Wright maps, which rely on
\emph{P}(\(\beta\)) = \(\tfrac12\), do not transfer to this family
without modification.

\includegraphics[width=1\textwidth,height=\textheight]{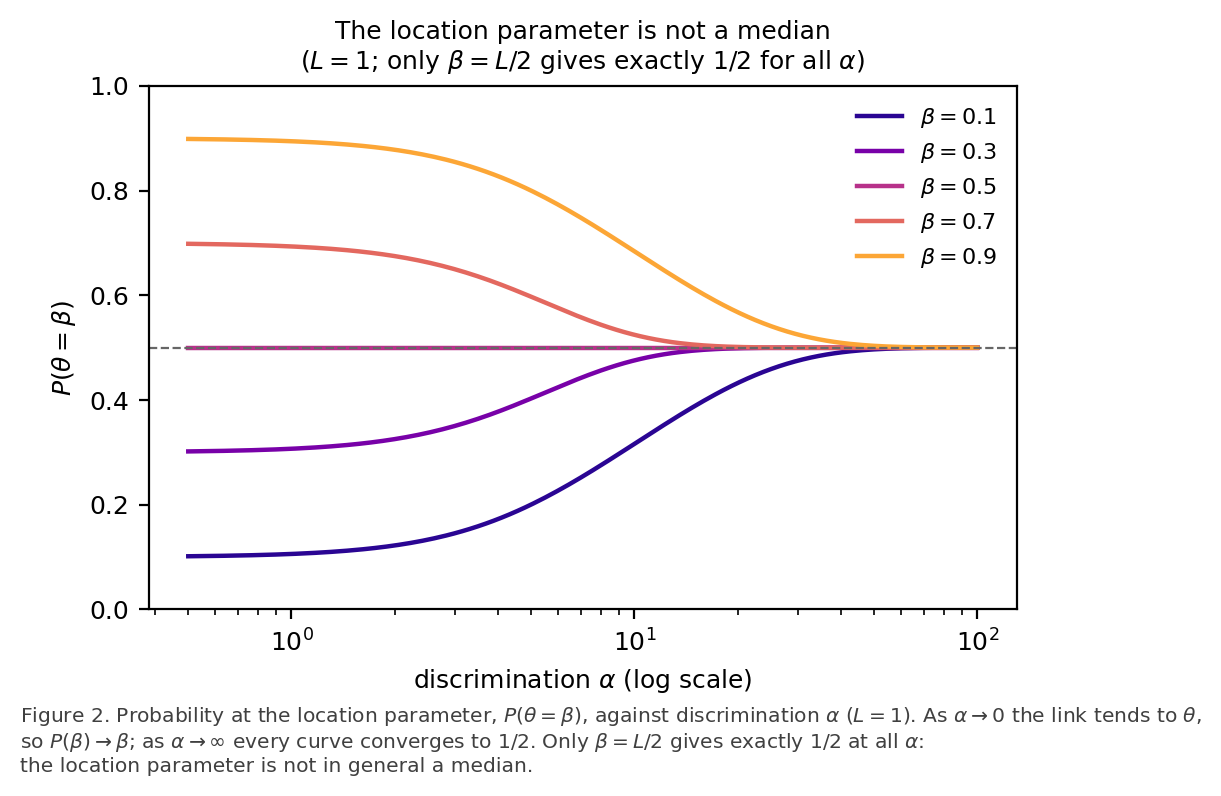}

\textbf{Proposition 2.6 (the weighted score is not sufficient).} The
statistic \emph{T}(\textbf{y}) = \(\Sigma\)\(_{j}\)
\(\alpha\)\(_{j}\)\emph{y}\(_{j}\) is not sufficient for \(\theta\).

\emph{Proof.} Sufficiency would require the likelihood ratio between any
two response patterns with equal \emph{T} to be constant in \(\theta\).
With \emph{J} = 20 items (\(\alpha\) = 5 for items 1--10, \(\alpha\) =
10 for items 11--20; \(\beta\) spread over {[}0.1, 1.0{]}) the patterns
\texttt{00000000001101000000} and \texttt{11111001000000000000} both
give \emph{T} = 30, yet their EAP estimates under a Beta(2, 2) prior
differ by 0.046. \(\square\)

The underlying reason is that log{[}\emph{P}/(1 \(-\) \emph{P}){]} is
not affine in \(\theta\) here, so the model is not an exponential family
in \(\theta\) with \emph{T} as its natural statistic --- in contrast to
the 2PL model, where it is. The family therefore uses response-pattern
information beyond the weighted score.

\begin{center}\rule{0.5\linewidth}{0.5pt}\end{center}

\hypertarget{numerical-form}{%
\subsection{2.6 Numerical form}\label{numerical-form}}

Evaluated as printed in (2.2), the link is stable across the parameter
range the source article considers. A direct transcription of Eq. (13)
in double precision reproduces the analytic value to approximately
10\(^{-}\)\(^{1}\)\(^{4}\) for \(\alpha\) as small as
10\(^{-}\)\(^{1}\)\(^{3}\), and departs from it only at \(\alpha\) =
10\(^{-}\)\(^{1}\)\(^{5}\). The single genuine failure is \(\alpha\) = 0
exactly, where 1 \(-\) \emph{e}\(^{0}\) = 0 gives 0/0; this is the
removable case treated by Proposition 2.4.

For the unbounded member \(\alpha\)\(\theta\) is not confined and
overflow becomes relevant. We evaluate (2.5) in log space,

log \emph{P}\(_{\infty }\) = logexpm1(\(\alpha\)\(\theta\)) \(-\)
\(\alpha\)\(\beta\) \(-\) softplus(\(\alpha\)(\(\theta\) \(-\)
\(\beta\))),

with logexpm1(\emph{u}) = \emph{u} + log1p(\(-\)\emph{e}\(^{-u}\)) for
large \emph{u}. This remains finite for \(\alpha\)\(\theta\) up to 5
\(\times\) 10\(^{4}\).

The literal transcription of Eqs. (12)--(14) is retained as a
correctness oracle: any optimized implementation is pinned by a test
asserting agreement with it to 10\(^{-}\)\(^{1}\)\(^{2}\) over
\(\alpha\) \(\in\) {[}0.1, 100{]}.

\begin{center}\rule{0.5\linewidth}{0.5pt}\end{center}

\hypertarget{corrections-to-the-source-article}{%
\subsection{2.7 Corrections to the source
article}\label{corrections-to-the-source-article}}

Three discrepancies were found while reproducing the published
derivation and results. None affects the equations as printed; two
concern the surrounding text and one concerns a table.

\textbf{(C1) The \(\alpha\) \(\to\) 0 limit (p.~6).} The text states
that Eq. (12) converges to \(\theta\) + \(\gamma\). By Proposition 2.4
the limit implied by the equation is \(\gamma\) + (1 \(-\)
\(\gamma\))\(\theta\); the two agree only when \(\gamma\) = 0 or
\(\theta\) = 0. \emph{(At \(\alpha\) = 10\(^{-}\)\(^{3}\), \(\theta\) =
0.5, \(\gamma\) = 0.2 the computed value is 0.600000, matching
\(\gamma\) + (1 \(-\) \(\gamma\))\(\theta\) = 0.600 rather than
\(\theta\) + \(\gamma\) = 0.700.)}

\textbf{(C2) The instability attributed to \(\alpha\) (\S{}3.1, Table
3).} The article concludes from RMSE(\(\alpha\)) = 0.962 versus
RMSE(\(\beta\)) = 0.050 that \(\alpha\) is estimated less stably than
\(\beta\) and \(\theta\). The two numbers are not commensurable. Root
mean squared error carries the units of the parameter, and \(\beta\) and
\(\theta\) are confined to {[}0, 1{]} whereas the generating values of
\(\alpha\) are 5 and 10; the ratio 0.962/0.050 is therefore a statement
about the parameters' scales, not about their estimability. The point is
not that the correct relative comparison is some particular number ---
as we show in \S{}3.2, that depends on how a relative error is defined,
and different reasonable definitions order \(\alpha\) and \(\beta\)
differently --- but that \textbf{no comparison of raw RMSE across
parameters on different scales licenses a conclusion about relative
stability}. On the one scale-free index the article itself reports, mean
relative bias, the three parameters are comparable (\(- 0.040,\)
\(- 0.032,\) 0.021). Our 20-replication reanalysis supplies the missing
counterpart from the other side of the estimator: on the same index our
estimates likewise place the three parameters in one band (+0.077,
+0.041, +0.069; \S{}3.2).

\(\alpha\) \emph{is} the hardest of the three to recover, for reasons we
identify in \S{}3.2 and \S{}6, but that conclusion requires evidence the
article does not present.

\textbf{(C3) Column alignment in Table 5.} In the 3P CTM half of Table 5
the median column appears to be duplicated, displacing the remaining
columns rightward and truncating the 97.5th percentile. Read under the
printed headers, the standard deviation equals the median and the 97.5th
percentile falls below the mean; read with one column removed, the
entries are coherent (e.g.~\(\alpha\)\(_{1}\): mean 8.86, median 8.60,
SD 1.70, 2.5th percentile 6.52). The 3PL half is unaffected.

\begin{center}\rule{0.5\linewidth}{0.5pt}\end{center}

\hypertarget{estimation}{%
\section{3. Estimation}\label{estimation}}

\begin{center}\rule{0.5\linewidth}{0.5pt}\end{center}

\hypertarget{quadrature-without-truncation}{%
\subsection{3.1 Quadrature without
truncation}\label{quadrature-without-truncation}}

Marginal estimation requires integrals of the form \(\int\)
\emph{f}(\(\theta\)) \emph{p}(\(\theta\)) d\(\theta\). In item response
theory \(\theta\) is unbounded, so these are approximated by
Gauss--Hermite quadrature (Bock \& Aitkin, 1981) on a range the analyst
must choose; the customary \(\pm 4\) standard deviations discards mass
that is small but not zero, and the choice becomes consequential if the
calibration sample drifts away from the prior.

For the \emph{L} = 1 member of the family the support is {[}0, 1{]} and
this problem does not arise. We use Gauss--Legendre nodes rescaled to
the unit interval, weighted by the Beta(\emph{a}, \emph{b}) prior and
renormalized:

\[\theta_k = \tfrac12(x_k+1), \qquad
w_k \propto \tfrac12 g_k \, \mathrm{Beta}(\theta_k; a, b),\]

where (\(x_k\), \(g_k\)) are the standard Legendre nodes and weights.
Since the Beta(2, 2) density is a quadratic polynomial, the rule
integrates the prior exactly: with as few as 21 nodes the recovered
prior mean and variance are 0.5000000000 and 0.0500000000 against the
analytic 0.5 and 0.05.

How many nodes are needed for the posterior is a separate question, and
the answer is fewer than one might expect. Taking a 61-node grid as
reference, EAP estimates for \emph{n} = 1000 examinees on \emph{J} = 20
items differ by

\begin{longtable}[]{@{}ll@{}}
\toprule\noalign{}
nodes & max \textbar{}\(\Delta\)EAP\textbar{} vs 61 \\
\midrule\noalign{}
\endhead
\bottomrule\noalign{}
\endlastfoot
11 & 1.7 \(\times\) 10\(^{-}\)\(^{3}\) \\
21 & 1.2 \(\times\) 10\(^{-}\)\(^{8}\) \\
41 & 4.3 \(\times\) 10\(^{-}\)\(^{1}\)\(^{5}\) \\
101 & 4.0 \(\times\) 10\(^{-}\)\(^{1}\)\(^{5}\) \\
\end{longtable}

\textbf{Forty-one nodes reach the double-precision floor.} We use 41
throughout and report 61 only where earlier results were computed with
it. The grid is fixed once and reused: it does not depend on the sample,
the items, or the occasion.

\begin{center}\rule{0.5\linewidth}{0.5pt}\end{center}

\hypertarget{item-parameters-by-bayes-modal-em}{%
\subsection{3.2 Item parameters by Bayes modal
EM}\label{item-parameters-by-bayes-modal-em}}

Let \emph{Y} be the \emph{n} \(\times\) \emph{J} response matrix. With
the priors used by Choi (2022) --- \(\theta\) \(\sim\) Beta(2, 2), log
\(\alpha\) \(\sim\) N(\(\mu\), \(\sigma\)\(^{2}\)), \(\beta\) \(\sim\)
Beta(2, 2), \(\gamma\) \(\sim\) Beta(7, 25) --- the posterior mode of
the item parameters can be obtained by an EM algorithm (Dempster, Laird,
\& Rubin, 1977) in which the E-step is a quadrature sum and the M-step
separates by item.

\textbf{E-step.} With current item parameters, form the posterior over
the grid for each examinee, and from it the expected counts

\[N_k = \sum_i p_{ik}, \qquad r_{jk} = \sum_i p_{ik}\, y_{ij},\]

where \(p_{ik}\) is examinee \emph{i}'s posterior weight at node
\emph{k}.

\textbf{M-step.} For each item \emph{j} independently, maximize

\[\sum_k \left[ r_{jk}\log P_j(\theta_k) + (N_k - r_{jk})\log\big(1-P_j(\theta_k)\big) \right]
+ \log \pi(\alpha_j,\beta_j,\gamma_j)\]

over the unconstrained parameters (\(\log\alpha_j\),
\(\operatorname{logit}\beta_j\), \(\operatorname{logit}\gamma_j\)). The
reparameterization enforces \(\alpha\) \textgreater{} 0 and \(\beta\),
\(\gamma\) \(\in\) (0, 1) automatically, so no constrained optimizer is
required; each item is a two- or three-dimensional problem. This is
Bayes modal estimation in the sense of Mislevy (1986), carried out on
bounded support.

\textbf{Replication of the published simulation.} Choi (2022) generated
1000 examinees from Beta(2, 2) responding to 20 items (\(\alpha\) = 5
for the first ten, \(\alpha\) = 10 for the rest; \(\beta\) spread over
{[}0.1, 1.0{]}) and estimated by MCMC, reporting mean bias (MB), mean
relative bias (MRB), RMSE and the Pearson correlation with the
generating values (their Table 3). Repeating the design 20 times and
estimating by Bayes modal EM followed by EAP scoring gives, on the same
four statistics (mean \(\pm\) SD over replications; published single-run
values alongside):

\begin{longtable}[]{@{}
  >{\raggedright\arraybackslash}p{(\columnwidth - 16\tabcolsep) * \real{0.1111}}
  >{\raggedright\arraybackslash}p{(\columnwidth - 16\tabcolsep) * \real{0.1111}}
  >{\raggedright\arraybackslash}p{(\columnwidth - 16\tabcolsep) * \real{0.1111}}
  >{\raggedright\arraybackslash}p{(\columnwidth - 16\tabcolsep) * \real{0.1111}}
  >{\raggedright\arraybackslash}p{(\columnwidth - 16\tabcolsep) * \real{0.1111}}
  >{\raggedright\arraybackslash}p{(\columnwidth - 16\tabcolsep) * \real{0.1111}}
  >{\raggedright\arraybackslash}p{(\columnwidth - 16\tabcolsep) * \real{0.1111}}
  >{\raggedright\arraybackslash}p{(\columnwidth - 16\tabcolsep) * \real{0.1111}}
  >{\raggedright\arraybackslash}p{(\columnwidth - 16\tabcolsep) * \real{0.1111}}@{}}
\toprule\noalign{}
\begin{minipage}[b]{\linewidth}\raggedright
\end{minipage} & \begin{minipage}[b]{\linewidth}\raggedright
MB
\end{minipage} & \begin{minipage}[b]{\linewidth}\raggedright
publ.
\end{minipage} & \begin{minipage}[b]{\linewidth}\raggedright
MRB
\end{minipage} & \begin{minipage}[b]{\linewidth}\raggedright
publ.
\end{minipage} & \begin{minipage}[b]{\linewidth}\raggedright
RMSE
\end{minipage} & \begin{minipage}[b]{\linewidth}\raggedright
publ.
\end{minipage} & \begin{minipage}[b]{\linewidth}\raggedright
\emph{r}
\end{minipage} & \begin{minipage}[b]{\linewidth}\raggedright
publ.
\end{minipage} \\
\midrule\noalign{}
\endhead
\bottomrule\noalign{}
\endlastfoot
\(\alpha\) & +0.527 \(\pm\) 0.235 & \(- 0.087\) & +0.077 \(\pm\) 0.031 &
\(- 0.040\) & 1.086 \(\pm\) 0.253 & 0.962 & 0.944 \(\pm\) 0.021 &
0.968 \\
\(\beta\) & \(- 0.009\) \(\pm\) 0.010 & \(- 0.002\) & +0.041 \(\pm\)
0.039 & \(- 0.032\) & 0.049 \(\pm\) 0.008 & 0.050 & 0.991 \(\pm\) 0.004
& 0.987 \\
\(\theta\) & +0.000 \(\pm\) 0.007 & \(- 0.004\) & +0.069 \(\pm\) 0.025 &
0.021 & 0.070 \(\pm\) 0.002 & 0.071 & 0.951 \(\pm\) 0.003 & 0.947 \\
\end{longtable}

For \(\beta\) and \(\theta\) the published values fall essentially on
our means, on all four statistics. For \(\alpha\) the RMSE and
correlation agree with the published run to within sampling variation,
while the biases carry the opposite sign (+0.527 here against
\(- 0.087\) there). The sign is a property of the estimator and prior
--- a posterior mode under log \(\alpha\) \(\sim\) N(log 10, 1) against
posterior medians from a sampler under a different \(\alpha\) prior ---
and we flag it rather than reconcile it; the comparison that survives
the change of estimator is the scale-invariant one below.

Two cautions apply to \(\alpha\), and the table now makes the first of
them precise. \textbf{On the scale-invariant MRB, the three parameters
are comparable}: +0.077 for \(\alpha\) against +0.041 for \(\beta\) and
+0.069 for \(\theta\). On RMSE, \(\alpha\) appears an order of magnitude
worse --- but RMSE for \(\alpha\) is measured in units of \(\alpha\),
and dividing by the generating value gives 17.0\% for the \(\alpha\) = 5
items and 12.2\% for the \(\alpha\) = 10 items, while the same operation
applied item-by-item to \(\beta\) gives 21.6\%, because \(\beta\) = 0.1
is a small denominator; dividing \(\beta\)'s RMSE by the mean generating
value instead gives 8.9\%. \textbf{\(\alpha\) and \(\beta\) change
places depending on the convention.} The inference drawn in the source
article from comparing raw RMSEs across parameters is taken up as
correction C2 in \S{}2.7; the numbers here are its empirical side.

The second caution is sharper. \textbf{The sampling standard deviation
of RMSE(\(\alpha\)) across replications is 0.253, twice the difference
between the published value and our mean}, so a single-replication
comparison of \(\alpha\) carries no information. The published value
lies comfortably inside our range, and any claim about \(\alpha\) ---
ours or the source article's --- must rest on repeated samples. This is
one reason we report every result in this paper with its replication
standard deviation.

\textbf{How well \(\alpha\) is recovered depends on where it is.}
Sweeping the generating value with the design otherwise fixed (\emph{n}
= 1000, \emph{J} = 20, all items sharing one \(\alpha\), prior centred
at the generating value; 20 replications each):

\begin{longtable}[]{@{}llll@{}}
\toprule\noalign{}
generating \(\alpha\) & RMSE & RMSE / \(\alpha\) & MRB \\
\midrule\noalign{}
\endhead
\bottomrule\noalign{}
\endlastfoot
2 & 1.023 \(\pm\) 0.107 & 51.2\% \(\pm\) 5.3\% & \(- 0.020\) \(\pm\)
0.101 \\
5 & 0.722 \(\pm\) 0.067 & 14.4\% \(\pm\) 1.3\% & +0.020 \(\pm\) 0.037 \\
10 & 0.977 \(\pm\) 0.188 & 9.8\% \(\pm\) 1.9\% & +0.006 \(\pm\) 0.023 \\
20 & 2.207 \(\pm\) 0.411 & 11.0\% \(\pm\) 2.1\% & +0.003 \(\pm\)
0.027 \\
40 & 7.917 \(\pm\) 2.459 & 19.8\% \(\pm\) 6.1\% & +0.032 \(\pm\)
0.035 \\
\end{longtable}

Relative recovery error is \textbf{U-shaped with its minimum around
\(\alpha\) = 10--20}. The mechanism is visible in the link function: at
\(\alpha\) = 2 the curve is nearly flat and the responses carry little
information about its slope, while at \(\alpha\) = 40 the curve is
nearly a step and every \(\alpha\) above the threshold fits the data
almost equally well. Absolute RMSE, by contrast, grows roughly in
proportion to \(\alpha\) --- another instance of the scale effect above.
For item writers the sweep doubles as a design guideline:
discriminations in the \(\alpha\) = 10--20 range are the ones this
design estimates best.

Convergence is fast: 88 iterations to a tolerance of 10\(^{-}\)\(^{6}\)
on the largest parameter change, in 1.4--2.8 s. The marginal
log-likelihood increases monotonically to within the tolerance of the
inner optimizer (maximum observed decrease 1.5 \(\times\)
10\(^{-}\)\(^{4}\) over 87 iterations, all in the final stages).

\begin{center}\rule{0.5\linewidth}{0.5pt}\end{center}

\hypertarget{person-scoring-eap-map-mle}{%
\subsection{3.3 Person scoring: EAP, MAP,
MLE}\label{person-scoring-eap-map-mle}}

Given item parameters, three point estimators are available --- the
posterior mean (EAP; Bock \& Mislevy, 1982), the posterior mode (MAP),
and the maximum likelihood estimate (MLE). Their accuracy is nearly
identical; what separates them is behaviour at the boundaries and the
amount of shrinkage.

Over 20 replications of a 3P design (\emph{n} = 2000, \emph{J} = 20,
\(\gamma\) = 0.2):

\begin{longtable}[]{@{}lll@{}}
\toprule\noalign{}
& RMSE (mean \(\pm\) SD) & boundary solutions \\
\midrule\noalign{}
\endhead
\bottomrule\noalign{}
\endlastfoot
EAP & 0.0838 \(\pm\) 0.0015 & 0 \\
MAP & 0.0846 \(\pm\) 0.0014 & 0 \\
MLE & 0.0927 \(\pm\) 0.0012 & 5.43\% \(\pm\) 0.53\% \\
observed proportion correct & 0.1363 \(\pm\) 0.0024 & --- \\
\end{longtable}

MAP is more accurate than MLE in every one of the 20 replications, and
MAP produced no boundary solution in any of them. Shrinkage is ordered
EAP \textgreater{} MAP \textgreater{} MLE: in the lowest and highest
quintiles of true \(\theta\) the mean biases are +0.053 / \(- 0.036\)
for EAP, +0.047 / \(- 0.019\) for MAP, and approximately zero for MLE.

\textbf{Boundary solutions are more dangerous here than in IRT.} In a
logistic IRT model an examinee with a perfect score receives
\(\hat{\theta}\) = +\(\infty\), and the failure is self-announcing. In
the bounded family the same examinee receives \(\hat{\theta}\) = 1, a
finite and interpretable value that Table 1 of the source article
glosses as the level at which \emph{all} tasks are performed
successfully. In our simulation the perfect scorers had true \(\theta\)
between 0.812 and 0.990 (mean 0.886); MLE reported 1.000 for all of
them, EAP 0.932 and MAP 0.957.

The rate depends on test length and on the model:

\begin{longtable}[]{@{}lll@{}}
\toprule\noalign{}
\emph{J} & 2P & 3P \\
\midrule\noalign{}
\endhead
\bottomrule\noalign{}
\endlastfoot
5 & 14.5\% & 23.5\% \\
10 & 8.1\% & 9.4\% \\
20 & 4.8\% & 5.7\% \\
40 & 1.7\% & 2.4\% \\
LSAT (\emph{J} = 5, 3P, empirical) & --- & \textbf{32.3\%} \\
\end{longtable}

\textbf{In the three-parameter model the character of the failure also
changes.} With \emph{J} = 20 and \(\gamma\) = 0.2, of the 103 boundary
solutions, 51 were at \(\hat{\theta}\) = 0 --- but only one examinee had
answered every item incorrectly. In the two-parameter model \emph{P}(0)
= 0 forces \(\hat{\theta}\) \textgreater{} 0 as soon as a single item is
answered correctly; with \(\gamma\) \textgreater{} 0 the probability of
a wrong answer is capped at 1 \(-\) \(\gamma\) even at \(\theta\) = 0,
so the likelihood flattens and unremarkable response patterns are pushed
to the boundary. \textbf{The extreme-pattern signal that allows boundary
cases to be screened in the 2P model is therefore not available in the
3P model.}

Guessing also degrades the observed proportion correct as an estimator:
its RMSE rises from 0.087 (2P) to 0.136 (3P), with mean bias +0.101,
because chance success inflates low-ability scores systematically.

\textbf{Recommendation.} Report \textbf{MAP as the point estimate and
the EAP posterior standard deviation as the uncertainty}. The two share
the same quadrature grid, so the pair costs nothing beyond either alone.
MLE is retained as a diagnostic and its boundary solutions are flagged
rather than reported as trait levels.

A side observation supports a recommendation made later (\S{}6): fixing
\(\gamma\) at its true value rather than estimating it changes person
scoring almost not at all (EAP RMSE 0.0825 versus 0.0831). The
instability of \(\gamma\) documented in \S{}6 is a problem for item
interpretation, not for scoring.

\begin{center}\rule{0.5\linewidth}{0.5pt}\end{center}

\hypertarget{computational-cost}{%
\subsection{3.4 Computational cost}\label{computational-cost}}

The source article estimates by MCMC in WinBUGS (Lunn, Thomas, Best, \&
Spiegelhalter, 2000). Because ``MCMC'' names a family of algorithms
rather than one, a speed comparison is meaningful only if the sampler is
specified and shown to have converged. We therefore report two samplers,
both targeting the same posterior as the EM algorithm.

\textbf{Sampler A --- random-walk Metropolis-within-Gibbs} (Metropolis
et al., 1953; Hastings, 1970). \(\theta\) is updated as \emph{n}
conditionally independent scalar draws (vectorized); each \(\alpha_j\)
is updated on the log scale with a Jacobian correction; each \(\beta_j\)
is updated on the natural scale with rejection outside (0, 1). Three
chains, 2000 burn-in and 3000 retained draws, matching the article's
simulation settings. Two variants: fixed step sizes, and Robbins--Monro
adaptation (Robbins \& Monro, 1951) during burn-in targeting an
acceptance rate of 0.44 (Gelman, Roberts, \& Gilks, 1996), with steps
frozen thereafter.

\textbf{Sampler B --- NUTS (Hoffman \& Gelman, 2014).} Implemented in
NumPyro (Phan, Pradhan, \& Jankowiak, 2019) with 64-bit precision, 2
chains, 400 warm-up and 400 retained draws, target acceptance 0.8.

On identical data (\emph{n} = 1000, \emph{J} = 20, 2P) and identical
priors:

\begin{longtable}[]{@{}lllll@{}}
\toprule\noalign{}
& wall clock & max \(\hat{R}\) & ESS/s & \(\theta\) RMSE \\
\midrule\noalign{}
\endhead
\bottomrule\noalign{}
\endlastfoot
\textbf{Bayes modal EM} & \textbf{1.4--2.8 s} & --- & --- & 0.0685 \\
MH, fixed step & 50.6 s (\textbf{35\(\times\)}) & 1.015 & 7.9 &
0.0685 \\
MH, adapted & 49.2 s (34\(\times\)) & 1.034 & 7.6 & 0.0683 \\
\textbf{NUTS} & 202.2 s (\textbf{73\(\times\)}) & 1.008 & 1.0--4.1 &
0.0682 \\
\end{longtable}

All three agree with each other and with the truth: correlations between
EM and NUTS estimates are 0.9987 (\(\alpha\)), 0.9997 (\(\beta\)) and
0.9999 (\(\theta\)), and the RMSE against the generating values is the
same to three decimal places. \textbf{Speed is not being bought with
accuracy.}

Two objections are worth pre-empting. First, that the random-walk
sampler was poorly tuned: adaptation moved the acceptance rates onto the
0.44 target (0.660/0.513/0.362 \(\to\) 0.439/0.443/0.435) without
improving efficiency (ESS/s 7.9 \(\to\) 7.6), so tuning is not the
binding constraint. Second, that a gradient-based sampler would close
the gap: it does not, and in fact widens it. The reason is structural.
Conditional on item parameters the \(\theta_i\) are independent, so a
Metropolis scheme updates all 1000 of them in one vectorized pass,
whereas NUTS computes a gradient with respect to all 1040 parameters and
takes several leapfrog steps per draw. The regime in which NUTS excels
--- high-dimensional posteriors with strong dependence --- is not this
one.

We state the claim with its scope: \textbf{against a converged MCMC
sampler of the kind the source article used, Bayes modal estimation is
35 times faster and gives the same answer; against NUTS, 73 times.} This
is not a claim against IRT, which has had marginal maximum likelihood
for forty years (Bock \& Aitkin, 1981).

\emph{Caveat.} NUTS was run with fewer total draws than the Metropolis
sampler (1600 versus 9000) for reasons of compute budget. \(\hat{R}\)
\(\le\) 1.008 indicates convergence, but the effective sample sizes are
not directly comparable in absolute terms; ESS per second, which is
invariant to run length, is the fair comparison and still favours the
Metropolis sampler.

\textbf{Adaptive administration.} The cost that matters for a response
loop is not calibration but the update after a single item. With the
grid and the item-by-node log-probability matrix precomputed once,
incorporating one response and recomputing the EAP requires a vector
addition, an exponentiation and a dot product over 41 nodes:

\begin{quote}
mean 4.7 \(\mu\)s, median 4.5 \(\mu\)s, 95th percentile 4.9 \(\mu\)s
(2000 simulated examinees \(\times\) 20 items; three timed passes pooled
after a warm-up pass) --- about 211,000 updates per second.
\end{quote}

Figure 3 shows the distribution: tight around the median, with a thin
upper tail folded into the last bin.

\includegraphics[width=1\textwidth,height=\textheight]{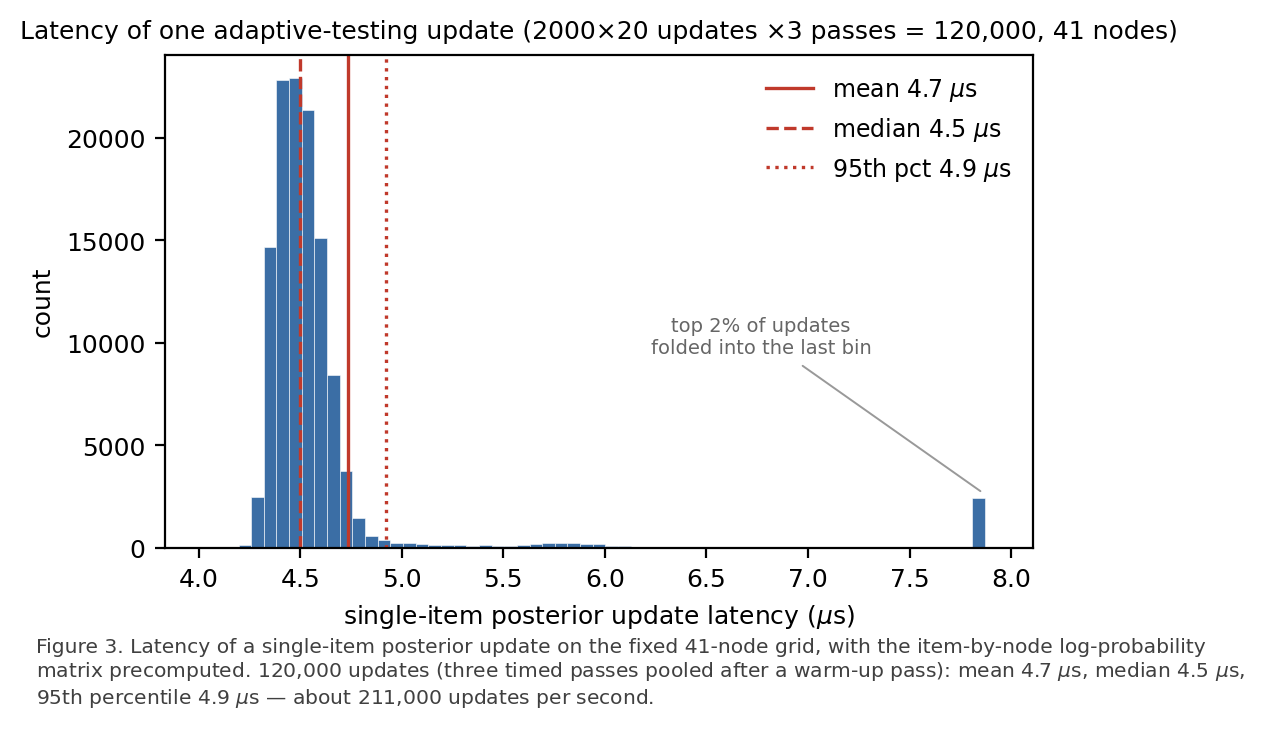}

Because the grid is fixed on {[}0, 1{]} independently of sample and
occasion, the precomputed matrix is valid for the lifetime of the item
bank. This is what places the model inside the response loop rather than
after it.

\begin{center}\rule{0.5\linewidth}{0.5pt}\end{center}

\hypertarget{information-and-what-it-does-at-the-boundaries}{%
\subsection{3.5 Information, and what it does at the
boundaries}\label{information-and-what-it-does-at-the-boundaries}}

The source article lists item and test information functions among its
suggestions for future work. The concepts are classical (Birnbaum,
1968); for this family they follow directly from Proposition 2.3.

Writing \emph{s}(\(\theta\)) = \(\sigma\)(\(\alpha\)(\(\theta\) \(-\)
\(\beta\))) and \emph{D} = \(\sigma\)(\(\alpha\)(\emph{L} \(-\)
\(\beta\))) \(-\) \(\sigma\)(\(- \alpha\)\(\beta\)), the derivative is
\emph{P\('\)} = (\(\alpha\)/\emph{D})\(\,\cdot\,\)\emph{s}(1 \(-\)
\emph{s}), and for a binary item

\[I(\theta) \;=\; \frac{[P'(\theta)]^{2}}{P(\theta)\,[1-P(\theta)]}
\;=\; \frac{(\alpha/D)^{2}\,s^{2}(1-s)^{2}}{P(1-P)} ,
\tag{3.1}\]

with \emph{I}\(_{3}\) = (1 \(-\)
\(\gamma\))\(^{2}\)(\emph{P}\(_{2}\)\('\))\(^{2}\)/{[}\emph{P}\(_{3}\)(1
\(-\) \emph{P}\(_{3}\)){]} in the three-parameter case. The test
information function is the sum over items and the standard error is
\emph{I}\(^{-\tfrac12}\).

\textbf{The numerator is bounded.} \emph{s}(1 \(-\) \emph{s}) \(\le\)
\(\tfrac14\), so \emph{P\('\)} \(\le\) \(\alpha\)/(4\emph{D})
everywhere. Any divergence of (3.1) therefore comes from the denominator
alone, and the question reduces to how fast \emph{P}(1 \(-\) \emph{P})
approaches zero at the edges of the support.

\textbf{Proposition 3.1 (boundary rate).} As \(\theta\) \(\to\)
\emph{L},

\[I(\theta) \;\sim\; \frac{P'(L)}{L-\theta} ,\]

and for the two-parameter model as \(\theta\) \(\to\) 0,
\emph{I}(\(\theta\)) \(\sim\) \emph{P}\('\)(0)/\(\theta\). In the
three-parameter model \emph{I}(0) is finite.

\emph{Proof.} Near \(\theta\) = \emph{L}, 1 \(-\) \emph{P} =
{[}\(\sigma\)(\(\alpha\)(\emph{L} \(-\) \(\beta\))) \(-\)
\emph{s}(\(\theta\)){]}/\emph{D} = \emph{P}\('\)(\emph{L})(\emph{L}
\(-\) \(\theta\)) + \emph{O}((\emph{L} \(-\) \(\theta\))\(^{2}\)), while
\emph{P} \(\to\) 1 and \emph{P}\('\) \(\to\) \emph{P}\('\)(\emph{L}).
Substituting into (3.1) gives \emph{I} \(\sim\)
\emph{P}\('\)(\emph{L})\(^{2}\)/{[}\emph{P}\('\)(\emph{L})(\emph{L}
\(-\) \(\theta\)){]}. The statement at \(\theta\) = 0 for \emph{L} = 1
follows symmetrically from \emph{P} = \emph{P}\('\)(0)\(\theta\) +
\emph{O}(\(\theta\)\(^{2}\)) and 1 \(-\) \emph{P} \(\to\) 1. For the
three-parameter model \emph{P}\(_{3}\)(0) = \(\gamma\) \textgreater{} 0,
so the denominator is bounded away from zero and \emph{I}\(_{3}\)(0) =
(1 \(-\)
\(\gamma\))\(^{2}\)\emph{P}\(_{2}\)\('\)(0)\(^{2}\)/{[}\(\gamma\)(1
\(-\) \(\gamma\)){]} is finite. \(\square\)

\emph{(Numerical check: (L \(-\) \(\theta\))\(\,\cdot\,\)I(\(\theta\))
at \(\theta\) = 1 \(-\) 10\(^{-}\)\(^{7}\) equals P\('\)(1) to six
decimal places for \(\alpha\) \(\in\) \{5, 10, 25\}; likewise
\(\theta\)\(\,\cdot\,\)I(\(\theta\)) \(\to\) P\('\)(0). An earlier
finite-difference estimate of the exponent gave 0.96--0.99; the exact
value is 1.)}

Three consequences follow.

\textbf{(i) The lower asymptote is a regularizer.} \(\gamma\) removes
the divergence at \(\theta\) = 0 but cannot touch the one at \(\theta\)
= \emph{L}, because \emph{P}(\emph{L}) = 1 regardless of \(\gamma\).
With \(\alpha\) = 10 and \(\beta\) = 0.5, \emph{I} at \(\theta\) =
10\(^{-}\)\(^{4}\) is 674.8 in the two-parameter model and 0.02 in the
three-parameter model; at \(\theta\) = 1 \(-\) 10\(^{-}\)\(^{4}\) it is
674.8 and 539.9 respectively.

\textbf{(ii) Total information is infinite.} Since the rate is exactly
\((L-\theta)^{-1}\), the integral of \emph{I} over the support diverges
logarithmically. Any summary that integrates information over the trait
continuum is therefore ill-defined for this family, and the standard
error \emph{I}\(^{-\tfrac12}\) tends to zero at the mastery boundary ---
an artefact of pinning \emph{P}(\emph{L}) = 1, not a statement about
precision. This is the reason we report the posterior standard deviation
instead (\S{}3.3).

\textbf{(iii) The pathology does not reach practice, provided the
interim estimate is not the MLE.} In an adaptive test the information
function is evaluated at the current estimate of \(\theta\), and an EAP
or MAP estimate never approaches the boundary closely enough for the
divergence to matter. In a simulated adaptive test (bank of 60 items, 20
administered, \emph{n} = 400, 3P):

\begin{longtable}[]{@{}
  >{\raggedright\arraybackslash}p{(\columnwidth - 8\tabcolsep) * \real{0.2000}}
  >{\raggedright\arraybackslash}p{(\columnwidth - 8\tabcolsep) * \real{0.2000}}
  >{\raggedright\arraybackslash}p{(\columnwidth - 8\tabcolsep) * \real{0.2000}}
  >{\raggedright\arraybackslash}p{(\columnwidth - 8\tabcolsep) * \real{0.2000}}
  >{\raggedright\arraybackslash}p{(\columnwidth - 8\tabcolsep) * \real{0.2000}}@{}}
\toprule\noalign{}
\begin{minipage}[b]{\linewidth}\raggedright
interim estimate
\end{minipage} & \begin{minipage}[b]{\linewidth}\raggedright
information
\end{minipage} & \begin{minipage}[b]{\linewidth}\raggedright
overall RMSE
\end{minipage} & \begin{minipage}[b]{\linewidth}\raggedright
top 20\%
\end{minipage} & \begin{minipage}[b]{\linewidth}\raggedright
boundary estimates
\end{minipage} \\
\midrule\noalign{}
\endhead
\bottomrule\noalign{}
\endlastfoot
EAP & as is & 0.0569 & 0.0511 & 0 \\
EAP & truncated to {[}0.02, 0.98{]} & 0.0569 & 0.0511 & 0 \\
MLE & as is & 0.0575 & 0.0576 & 1,036 \\
MLE & truncated & \textbf{0.0989} & \textbf{0.1483} & \textbf{5,417} \\
\end{longtable}

With EAP the truncation never activates: the two rows agree to four
decimal places. With MLE it is actively harmful --- an examinee driven
to the boundary has zero truncated information at every item, item
selection becomes effectively random, and the number of boundary
estimates rises fivefold, tripling the RMSE among high performers.
\textbf{The remedy is not to modify the information function but to
avoid the maximum likelihood estimate as the interim estimate}, which is
the same recommendation reached in \S{}3.3 on independent grounds.

Finally, information need not be used for item selection at all.
Choosing the item that minimizes the expected posterior variance (MEPV;
Owen, 1975; van der Linden, 1998) is slightly more accurate (RMSE 0.0528
versus 0.0569 for maximum information, against 0.0776 for random
selection) and is immune to the divergence by construction, since it
never evaluates \emph{I}. It costs more, because each candidate item
requires the posterior to be updated twice, but not enough to matter:
over a 60-item bank the full cycle of posterior update, item selection
and EAP recomputation takes

\begin{quote}
22.9 \(\mu\)s (median 21.9, 95th percentile 32.9) for MEPV against 12.1
\(\mu\)s (median 11.5, 95th percentile 13.8) for maximum information ---
\end{quote}

a factor of 1.9, still about 40,000 selections per second.

\begin{center}\rule{0.5\linewidth}{0.5pt}\end{center}

\hypertarget{what-has-been-established-and-what-has-not}{%
\subsection{3.6 What has been established, and what has
not}\label{what-has-been-established-and-what-has-not}}

Bounded support removes the estimation obstacle. Quadrature is exact in
the prior and at machine precision with 41 nodes; the item parameters
follow from a deterministic EM whose answers match MCMC at one to two
orders of magnitude lower cost; and the fixed grid makes a single-item
update a microsecond operation valid for the lifetime of the item bank.
The one pathology inherited from the anchoring --- divergent information
at the mastery boundary --- is harmless with a posterior mean or mode
and harmful only in combination with the maximum likelihood estimate,
which we recommend against on two other grounds as well.

None of this says anything about whether the model \emph{measures}
better than item response theory. Speed and interpretability are not
accuracy, and a reader entitled to ask what the bounded scale buys
psychometrically has not yet been answered. Section 4 takes up that
question directly, and the answer is mostly negative.

\begin{center}\rule{0.5\linewidth}{0.5pt}\end{center}

\hypertarget{what-the-bounded-scale-buys-and-what-it-does-not}{%
\section{4. What the Bounded Scale Buys, and What It Does
Not}\label{what-the-bounded-scale-buys-and-what-it-does-not}}

\begin{center}\rule{0.5\linewidth}{0.5pt}\end{center}

\hypertarget{is-the-family-a-reparameterization-of-irt}{%
\subsection{4.1 Is the family a reparameterization of
IRT?}\label{is-the-family-a-reparameterization-of-irt}}

The natural first suspicion is that nothing here is new: that
\emph{P\(_{L}\)} is the two-parameter logistic function seen through a
change of variable, in which case no comparison of measurement
properties could come out anything but a tie. The suspicion is half
right, and the half that is wrong turns out to matter.

Under a reparameterization \(\theta\) \(\mapsto\) \emph{g}(\(\theta\))
with \emph{g} strictly increasing, the two-parameter logistic response
function becomes \(\sigma\)(\emph{a\(_{j}\)}(\emph{g}(\(\theta\)) \(-\)
\emph{b\(_{j}\)})), whose logit is affine in \emph{g}(\(\theta\))
\textbf{with the same \emph{g} for every item} (Lord, 1975; Lord, 1980).
Consequently, if one model were a reparameterization of the other, the
logit of any one item's response function would be an affine function of
the logit of any other's. That is a testable condition.

The condition holds for some item pairs and fails for others, and the
pattern is exact rather than approximate.

\begin{longtable}[]{@{}lll@{}}
\toprule\noalign{}
item pair (\(\alpha\), \(\beta\)) & affine fit \emph{R}\(^{2}\) & max
residual \\
\midrule\noalign{}
\endhead
\bottomrule\noalign{}
\endlastfoot
(5, 0.3) vs (5, 0.7) --- \emph{same} \(\alpha\) & 1.000000 & 0.0000 \\
(5, 0.3) vs (10, 0.3) & 0.996368 & 0.703 \\
(5, 0.3) vs (15, 0.8) & 0.991309 & 1.531 \\
(2, 0.2) vs (20, 0.9) & 0.977615 & 3.173 \\
\emph{2PL control} & 1.000000 & 2 \(\times\)
10\(^{-}\)\(^{1}\)\(^{4}\) \\
\end{longtable}

The reason is the following decomposition, which also explains why
\(\beta\) never matters and \(\alpha\) always does.

\textbf{Proposition 4.1 (separation of \(\beta\)).} For the \emph{L} = 1
member,

\[\operatorname{logit} P(\theta;\alpha,\beta) \;=\; -\log A(\beta) \;+\; h(\theta;\alpha),
\qquad
h(\theta;\alpha) = -\log\!\left[\frac{1}{e^{\alpha\theta}-1} - \frac{1}{e^{\alpha}-1}\right],\]

where \emph{A}(\(\beta\)) = (1 +
\emph{e}\(^{-\alpha \beta }\))/\emph{C}(\(\beta\)) and
\emph{C}(\(\beta\)) is the normalizer of Eq. (2.4). The second term does
not involve \(\beta\).

For a general anchor \emph{L} the same argument gives \(\rho\) =
\(- 1\)/(\emph{e}\(^{\alpha L}\) \(-\) 1), so the separation holds
throughout the family. In the limit \emph{L} \(\to\) \(\infty\) the
constant vanishes and \emph{h} takes its simplest form,
\emph{h}\(_{\infty }\)(\(\theta\); \(\alpha\)) =
log(\emph{e}\(^{\alpha \theta }\) \(-\) 1) --- that is, \textbf{for HCTM
the logit is exactly \(\log(e^{\alpha\theta} - 1)\) shifted by a
function of \(\beta\) alone.}

\emph{Proof.} Write \emph{E} = \emph{e}\(^{\alpha \theta }\) \(-\) 1 and
\emph{P} = \emph{CE}/(1 + \emph{e}\(^{\alpha (\theta -\beta )}\)).
Substituting \emph{e}\(^{\alpha (\theta -\beta )}\) =
\emph{e}\(^{-\alpha \beta }\)(\emph{E} + 1) into 1 \(-\) \emph{P} and
collecting terms gives

\[\frac{1}{\text{odds}} \;=\; A(\beta)\left[\frac{1}{E} + \rho\right],
\qquad \rho = \frac{e^{-\alpha\beta}-C(\beta)}{1+e^{-\alpha\beta}} .\]

Substituting \emph{C}(\(\beta\)) = (1 +
\emph{e}\(^{\alpha (1-\beta )}\))/(\emph{e}\(^{\alpha }\) \(-\) 1) into
\(\rho\) and simplifying yields \(\rho\) =
\(- 1\)/(\emph{e}\(^{\alpha }\) \(-\) 1), which is free of \(\beta\).
Taking logarithms gives the statement. \(\square\)

\emph{(Numerically, \(\rho\) agrees with \(-1/(e^{\alpha} - 1)\) to
twelve decimal places across \(\alpha\) \(\in\) {[}0.5, 25{]} and
\(\beta\) \(\in\) {[}0.05, 0.95{]}; for HCTM the fitted \(\rho\) is 0 to
ten decimal places, and the logit regression between same-\(\alpha\)
items has slope 1.0000000000 with residual SD \(\le\) 3 \(\times\)
10\(^{-}\)\(^{1}\)\(^{2}\).)}

\textbf{Corollary.} Two items with the same \(\alpha\) have logits
differing by the constant log \emph{A}(\(\beta\)\(_{1}\)) \(-\) log
\emph{A}(\(\beta\)\(_{2}\)); their logit regression has slope exactly 1.
Two items with different \(\alpha\) do not, because
\emph{h}(\(\,\cdot\,\); \(\alpha\)) differs in shape.

It follows that \textbf{the one-parameter member --- in which \(\alpha\)
is fixed by fiat --- is a reparameterization of the one-parameter
logistic model (Rasch, 1960), with \emph{g}(\(\theta\)) =
\emph{h}(\(\theta\); \(\alpha\)) as the common transformation, and a tie
in measurement properties is guaranteed for it.} The two- and
three-parameter members are not reparameterizations: they are affine
transforms in \emph{probability} space, and affine transforms of a
response function alter \emph{P}(1 \(-\) \emph{P}) and \emph{P}\('\),
hence alter information.

So for 2P and 3P the family and item response theory are genuinely
different models, and whether one measures better than the other is an
empirical question rather than a tautology. We put it to three tests.
All three came out null.

\begin{center}\rule{0.5\linewidth}{0.5pt}\end{center}

\hypertarget{three-attempts-to-find-a-measurement-advantage}{%
\subsection{4.2 Three attempts to find a measurement
advantage}\label{three-attempts-to-find-a-measurement-advantage}}

\textbf{Test 1: discrimination at the ceiling.} If the bounded scale
helps anywhere, it should help where the unbounded scale is worst ---
among high performers on a test that is too easy for them. We generated
responses from a 2PL model (the choice favours item response theory),
held items fixed, and shifted the population mean upward until a quarter
of examinees answered every item correctly.

\begin{longtable}[]{@{}
  >{\raggedright\arraybackslash}p{(\columnwidth - 8\tabcolsep) * \real{0.2000}}
  >{\raggedright\arraybackslash}p{(\columnwidth - 8\tabcolsep) * \real{0.2000}}
  >{\raggedright\arraybackslash}p{(\columnwidth - 8\tabcolsep) * \real{0.2000}}
  >{\raggedright\arraybackslash}p{(\columnwidth - 8\tabcolsep) * \real{0.2000}}
  >{\raggedright\arraybackslash}p{(\columnwidth - 8\tabcolsep) * \real{0.2000}}@{}}
\toprule\noalign{}
\begin{minipage}[b]{\linewidth}\raggedright
population mean
\end{minipage} & \begin{minipage}[b]{\linewidth}\raggedright
perfect scores
\end{minipage} & \begin{minipage}[b]{\linewidth}\raggedright
top-20\% Spearman, IRT
\end{minipage} & \begin{minipage}[b]{\linewidth}\raggedright
CTM
\end{minipage} & \begin{minipage}[b]{\linewidth}\raggedright
distinct estimates in top 20\%, IRT / CTM
\end{minipage} \\
\midrule\noalign{}
\endhead
\bottomrule\noalign{}
\endlastfoot
0 & 0.2\% & 0.545 & 0.537 & 369 / 368 \\
2 & 7.4\% & 0.375 & 0.367 & 113 / 113 \\
3 & 24.7\% & 0.276 & 0.281 & 43.3 / 43.3 \\
\end{longtable}

The rank correlations differ by less than 0.01, and the count of
distinct estimates is identical. That second column is the informative
one: examinees who answer every item correctly share one response
pattern, and \textbf{no model can separate examinees whose data are
identical.} The ceiling effect is a loss of information in the data, not
a defect of the scale, and bounding the scale therefore cannot repair
it.

\textbf{Test 2: invariance of the estimate.} The same 200 reference
examinees, with fixed true abilities, were placed into populations of
increasing mean, the item bank held fixed and recalibrated each time.
Neither model keeps the estimate still: expressed as a fraction of each
model's own spread, moving the population from mean 0 to mean 3 shifts
it by \textbf{71\% for IRT and 69\% for CTM}. Since the result bears on
the interpretive claim rather than on measurement accuracy, we defer its
analysis to \S{}4.5.

\textbf{Test 3: growth tracking.} With items calibrated once and then
anchored, a cohort of 1500 learners was followed over four occasions
along a learning curve. Rank recovery at the final occasion was 0.8307
(CTM) against 0.8299 (IRT); recovery of individual gains was 0.8496
against 0.8549; the proportion of real growth that went undetected was
0.2\% in both.

\textbf{These are not the results we set out to obtain}, and we report
them because they bound what the rest of the paper may claim. The
contribution of the bounded scale is not that it measures better.

\begin{center}\rule{0.5\linewidth}{0.5pt}\end{center}

\hypertarget{a-methodological-caution-about-rank-correlation}{%
\subsection{4.3 A methodological caution about rank
correlation}\label{a-methodological-caution-about-rank-correlation}}

Before stating the caution in general terms, it helps to have the source
article's own instrument on the table. Table 4 of Choi (2022)
correlates, on the simulation design of \S{}3.2, the true trait with
three estimates of it --- the observed proportion correct (CP), the
two-parameter logistic estimate, and the bounded-model estimate.
Reproducing it across 20 replications, with EM-calibrated and EAP-scored
analogues of the article's MCMC medians:

\begin{longtable}[]{@{}
  >{\raggedright\arraybackslash}p{(\columnwidth - 8\tabcolsep) * \real{0.2000}}
  >{\raggedright\arraybackslash}p{(\columnwidth - 8\tabcolsep) * \real{0.2000}}
  >{\raggedright\arraybackslash}p{(\columnwidth - 8\tabcolsep) * \real{0.2000}}
  >{\raggedright\arraybackslash}p{(\columnwidth - 8\tabcolsep) * \real{0.2000}}
  >{\raggedright\arraybackslash}p{(\columnwidth - 8\tabcolsep) * \real{0.2000}}@{}}
\toprule\noalign{}
\begin{minipage}[b]{\linewidth}\raggedright
pair
\end{minipage} & \begin{minipage}[b]{\linewidth}\raggedright
Pearson (mean \(\pm\) SD)
\end{minipage} & \begin{minipage}[b]{\linewidth}\raggedright
publ.
\end{minipage} & \begin{minipage}[b]{\linewidth}\raggedright
Spearman (mean \(\pm\) SD)
\end{minipage} & \begin{minipage}[b]{\linewidth}\raggedright
publ.
\end{minipage} \\
\midrule\noalign{}
\endhead
\bottomrule\noalign{}
\endlastfoot
(True, CP) & 0.949 \(\pm\) 0.003 & 0.944 & 0.949 \(\pm\) 0.003 &
0.945 \\
(True, 2PL) & 0.947 \(\pm\) 0.003 & 0.944 & 0.951 \(\pm\) 0.003 &
0.947 \\
(True, CTM) & 0.951 \(\pm\) 0.003 & 0.947 & 0.951 \(\pm\) 0.003 &
0.947 \\
(CP, 2PL) & 0.993 \(\pm\) 0.001 & 0.993 & 0.997 \(\pm\) 0.001 & 0.996 \\
(CP, CTM) & 0.997 \(\pm\) 0.001 & 0.995 & 0.997 \(\pm\) 0.001 & 0.995 \\
(2PL, CTM) & 0.996 \(\pm\) 0.000 & 0.997 & 0.9999 \(\pm\) 0.0000 &
1.000 \\
\end{longtable}

The reproduction is clean: every entry falls within 0.005 of the
published value. What the table \emph{means} is another matter, and it
is uninformative in two distinct ways.

First, the entry usually singled out --- a rank correlation between the
two models' estimates reported as a perfect 1 --- is a rounding
artefact. At full precision the correlation is below 1 in \textbf{every
one} of the 20 replications (largest 0.999931, smallest 0.999797). This
is what Proposition 4.1 predicts: the design mixes \(\alpha\) = 5 and
\(\alpha\) = 10 items, the two models are therefore not monotone
transforms of one another, and small systematic rank disagreements must
exist. The perfect 1 in the published table is the three-decimal shadow
of a correlation that is provably not 1.

Second, and more consequentially, even the exact value answers no
question a practitioner asks. Agreement between two estimators is
compatible with both being accurate and with both being wrong, and
0.9999 against 1.0000 changes nothing for either reading. Cashing the
table out on a scale that has a true value --- the domain score defined
below, computed on the same replications --- settles it: the two models
are equivalent (RMSE 0.0733 \(\pm\) 0.0018 against 0.0734 \(\pm\)
0.0018) and both beat the observed proportion correct (0.0791 \(\pm\)
0.0020). On the source article's own simulation design, its Table 4,
made to say something checkable, says what \S{}4.2 found by other means:
the measurement null. (The condition-dependent differences of \S{}4.4
arise under a different generating model and a displaced population; on
the calibration design there are none.)

The general reason to distrust the index follows. Rank correlation is
invariant under any strictly increasing transformation of the estimates.
If two models order examinees identically --- as models fitted to the
same responses very often will --- their Spearman correlations with the
truth are identical \emph{by construction}, whatever the scales, priors
or quadrature grids. The index cannot see the things we were varying.

The symptom is visible in our own output: across conditions that
differed in quadrature range and in prior, Spearman correlations agreed
to four decimal places, repeatedly. That is not stability; it is
blindness.

Replacing it with a scale-free quantity that has a true value --- the
expected proportion correct over the whole item bank, the \emph{domain
score} of Bock, Thissen and Zimowski (1997) --- makes the comparison
answer to the data: differences appear where they exist (\S{}4.4), and
their absence becomes a finding rather than a blind spot (the table
above). Each model predicts it from its own item parameters and its own
posterior,

\[\widehat{DS}_i \;=\; \int \frac{1}{J}\sum_{j} P_j^{\text{model}}(\theta)\;
p(\theta \mid \mathbf{y}_i)\, d\theta ,\]

so the comparison uses no knowledge of the truth and is symmetric
between models.

\textbf{The source article's Tables 4 and 6 report correlations only},
and share this limitation. We recommend that comparisons of this kind
report domain-score error, the number of distinct estimates produced, or
credible-interval coverage alongside any correlation.

\begin{center}\rule{0.5\linewidth}{0.5pt}\end{center}

\hypertarget{where-differences-do-appear}{%
\subsection{4.4 Where differences do
appear}\label{where-differences-do-appear}}

With the domain score as criterion, and 20 replications of each
condition (responses generated from a 2PL model, \emph{n} = 1500,
\emph{J} = 25):

\begin{longtable}[]{@{}
  >{\raggedright\arraybackslash}p{(\columnwidth - 8\tabcolsep) * \real{0.2000}}
  >{\raggedright\arraybackslash}p{(\columnwidth - 8\tabcolsep) * \real{0.2000}}
  >{\raggedright\arraybackslash}p{(\columnwidth - 8\tabcolsep) * \real{0.2000}}
  >{\raggedright\arraybackslash}p{(\columnwidth - 8\tabcolsep) * \real{0.2000}}
  >{\raggedright\arraybackslash}p{(\columnwidth - 8\tabcolsep) * \real{0.2000}}@{}}
\toprule\noalign{}
\begin{minipage}[b]{\linewidth}\raggedright
condition
\end{minipage} & \begin{minipage}[b]{\linewidth}\raggedright
CTM
\end{minipage} & \begin{minipage}[b]{\linewidth}\raggedright
IRT
\end{minipage} & \begin{minipage}[b]{\linewidth}\raggedright
observed proportion correct
\end{minipage} & \begin{minipage}[b]{\linewidth}\raggedright
direction consistent
\end{minipage} \\
\midrule\noalign{}
\endhead
\bottomrule\noalign{}
\endlastfoot
full bank, population mean 0 & 0.0764 \(\pm\) 0.0015 & \textbf{0.0749}
\(\pm\) 0.0016 & 0.0831 \(\pm\) 0.0016 & 20/20 \\
12 easy items only & 0.1042 \(\pm\) 0.0018 & \textbf{0.1001} \(\pm\)
0.0017 & \textbf{0.2378} \(\pm\) 0.0025 & 20/20 \\
full bank, population mean 3 & \textbf{0.0684} \(\pm\) 0.0012 & 0.0941
\(\pm\) 0.0011 & 0.0542 & 20/20 \\
\end{longtable}

Two things follow.

\textbf{(a) The direction depends on the condition, and it is
consistent.} Under the calibration population and under partial
administration, item response theory is slightly ahead; under a
population displaced from the calibration sample, the bounded model is
ahead by 27\%. The differences are small --- 0.002 to 0.004 in the first
two rows --- but every one of the 20 replications agrees in sign, so
they are not sampling noise. (That the generating model here is the 2PL
is why a small consistent IRT edge appears in the first row at all; on
the CTM-generated design of \S{}4.3 the same comparison is a tie.) The
correct statement is not that the models tie but that \textbf{neither
dominates}, which is what \S{}4.1 leads one to expect of two models that
differ but not by much.

The mechanism in the third row is the shrinkage distance. A N(0, 1)
prior can pull an estimate arbitrarily far toward the calibration mean;
a Beta(2, 2) prior on {[}0, 1{]} can pull it at most to the middle of a
unit interval, and as \(\theta\) \(\to\) 1 the response probabilities
saturate so that errors in \(\theta\) cost little in domain score.

\textbf{(b) Model-based scoring earns its keep only under some
conditions.} When examinees are administered a subset of the bank, both
models beat the observed proportion correct by a factor of two (0.10
against 0.24) --- the raw score of an easy subset is simply not an
estimate of performance on the whole domain. But when the whole bank is
administered to a cohort near the ceiling, \textbf{the observed
proportion correct beats both models} (0.054 against 0.068 and 0.094):
there, shrinkage is a pure cost. Taken together, these two rows locate
the conditions under which a latent variable model is worth its
complexity --- when examinees receive different items, and not when a
fixed form is given to a homogeneous high-performing group.

\begin{center}\rule{0.5\linewidth}{0.5pt}\end{center}

\hypertarget{what-actually-holds-the-scale}{%
\subsection{4.5 What actually holds the
scale}\label{what-actually-holds-the-scale}}

The second test of \S{}4.2 left a question open. A fixed examinee's
estimate moved by 69\% of the score spread when the population around
them changed, even though the item bank was the same throughout. If the
trait is anchored to the task domain, as the construction asserts, that
should not happen --- and the natural next question is whether changing
the \emph{domain} produces a comparable effect.

We therefore ran the complementary experiment: hold the population fixed
and change the domain. A 40-item bank was split into its easier and
harder halves, and each subset was calibrated and scored separately.

\begin{longtable}[]{@{}
  >{\raggedright\arraybackslash}p{(\columnwidth - 4\tabcolsep) * \real{0.3333}}
  >{\raggedright\arraybackslash}p{(\columnwidth - 4\tabcolsep) * \real{0.3333}}
  >{\raggedright\arraybackslash}p{(\columnwidth - 4\tabcolsep) * \real{0.3333}}@{}}
\toprule\noalign{}
\begin{minipage}[b]{\linewidth}\raggedright
manipulation
\end{minipage} & \begin{minipage}[b]{\linewidth}\raggedright
held fixed
\end{minipage} & \begin{minipage}[b]{\linewidth}\raggedright
shift in \(\hat{\theta}\) (as \% of its own spread)
\end{minipage} \\
\midrule\noalign{}
\endhead
\bottomrule\noalign{}
\endlastfoot
population moves from mean 0 to 3 & prior, items & \textbf{69\%} \\
domain replaced by its easier or harder half & prior, population &
\textbf{\(\le\) 1.5\%} \\
\end{longtable}

The item difficulties estimated from the two halves agree to within
0.008 on shared items, and the mean of \(\hat{\theta}\) moves from 0.499
to 0.503 and 0.497 although the subsets have proportions correct of
0.720 and 0.276.

\textbf{The scale is held by the prior, not by the task domain.} Under
marginal maximum likelihood the \(\theta\) distribution is forced toward
Beta(2, 2); if the items are easy, the item difficulty estimates absorb
the easiness and the trait distribution returns to where the prior puts
it. The construction assigns 0 and 1 to the ignorance and mastery levels
of a domain, but the estimator re-anchors them to the reference
population.

This is not a defect peculiar to the family. Item response theory has
the same dependence and has always required linking or anchoring to
overcome it. What is peculiar is the \emph{direction} of the exposure.
\textbf{The source article anticipates instability from changes to the
task domain and does not discuss the population; we find the domain
harmless and the population consequential.} A practitioner who reads
\(\hat{\theta}\) = 0.62 as ``62\% of mastery'' is entitled to know that
the figure is 62\% of mastery \emph{as located by the reference
population used at calibration}, and that the reading transfers across
cohorts only if the item parameters are anchored.

This is the sense in which the origin is what must be secured, and the
sense in which current practice does not secure it.

One asymmetry survives this qualification, and it is not about accuracy.
Because \(\theta\) has an external referent, the bias of an estimate can
be stated as a number: in the growth simulation of \S{}4.2 the mean
estimate exceeded the mean true value by 0.156. In the item response
theory metric that quantity does not exist, because there is no value
the sample mean \emph{should} have taken --- the mean is 0 by
construction. What the bounded scale offers is therefore not a more
accurate estimate but an \textbf{auditable} one: a claim about the trait
that can be wrong in a stateable amount, and hence checked.

A second exposure is independent of the first. That the relation between
a trait metric and a number-correct metric is in general nonlinear is
classical: Lord (1953) showed that if the trait metric is taken to
provide equal units then the true-score units cannot be, and that no
single trait metric linearizes the score--trait regression for all tests
of the same trait at once. The requirement that a binomial,
domain-referenced interpretation hold exactly is likewise restrictive,
demanding equal item difficulties (van der Linden, 1979). What is
specific to the bounded construction is where its two boundary
conditions leave that relation: for every item \(P(0) = 0\) and
\(P(L) = 1\), which pin \(\theta\) and the expected proportion of the
domain to agree exactly at the two endpoints and pin nothing between
them. The expected proportion at a given \(\theta\) is the item-average
of the response functions, so it depends on how the difficulties are
distributed across the domain. In the LSAT data of \S{}6 the two diverge
substantially: under the three-parameter model an examinee at \(\theta\)
= 0.20 is expected to answer a proportion 0.55 of the items, and one at
\(\theta\) = 0.50, a proportion 0.78 (Appendix, ``\(\theta\) and the
expected proportion of the domain''). Where the domain proportion is the
quantity of interest it should be computed and reported directly as a
domain score (Bock, Thissen, \& Zimowski, 1997) rather than read off
\(\theta\). A full characterization of when the two coincide is left to
future work.

\emph{Remark.} The two boundary conditions fix the ends of the scale but
not the interval between them. The situation is the one that preceded
thermodynamic temperature, where two fixed points were agreed and what
lay at the midpoint was whatever the working substance made it (Chang,
2004); the parallel with psychometrics is not new (Choppin, 1985). What
plays the part of the working substance here is the distribution of item
difficulty across the domain.

It is worth saying where this leaves \(\theta\). Three quantities answer
the question ``how much of it does this person have'' on a {[}0, 1{]}
scale, and all three fix the same two ends: the proportion correct, the
percentile rank, and \(\theta\). They differ in what fills the interval.
The proportion correct is filled by the items administered, so the same
number means different things on an easy and a hard form. The percentile
rank is filled by the reference group, so the same person moves when the
cohort does. \(\theta\) is filled by the model together with the task
domain: the item parameters are modelled explicitly, so the coordinate
does not depend on which form was taken, and its origin is an attainable
level rather than the asymptote of Lord's (1953) restriction III. That
is a different position among the three, not a better one --- \S{}4.2
found no measurement advantage, and the exposure documented above runs
through the calibration population. What the position does offer is that
its two commitments are stated and can be checked.

Within a fixed domain and a calibrated bank the coordinate is
reproducible: the grid is fixed once (\S{}3.1), the three point
estimators agree to within 0.07 at twenty items, and the reported
posterior standard deviation tracks the actual error (\S{}3.3, \S{}5.4).
How far that stability extends --- across calibration samples, prior
specifications and test lengths --- is a question we set out
systematically elsewhere.

\begin{center}\rule{0.5\linewidth}{0.5pt}\end{center}

\hypertarget{sequential-updating-and-growth}{%
\section{5. Sequential Updating and
Growth}\label{sequential-updating-and-growth}}

\begin{center}\rule{0.5\linewidth}{0.5pt}\end{center}

\hypertarget{the-obvious-extension-and-why-it-fails}{%
\subsection{5.1 The obvious extension, and why it
fails}\label{the-obvious-extension-and-why-it-fails}}

Section 4.5 concluded that the trait scale is held by the prior rather
than by the task domain. Read as a statement about a single occasion,
that is an embarrassment. Read as a statement about a \emph{sequence} of
occasions, it is a design.

Growth is expressed here on the \(\theta\) scale; because \(\theta\) and
the expected proportion of the domain do not coincide in general
(\S{}4.5), a gain in \(\theta\) should not be read as a gain of the same
size in the proportion of tasks performed.

If the prior is what locates the scale, then the natural object to carry
between occasions is the prior itself. Bayesian updating supplies the
mechanism without further apparatus: the posterior at occasion \emph{t}
becomes the prior at \emph{t} + 1. The bounded support makes this
unusually cheap, because a posterior is a vector of weights on a grid
that never changes --- 41 numbers per learner, or, as we show below,
two.

The obvious version of this does not work.

\begin{longtable}[]{@{}
  >{\raggedright\arraybackslash}p{(\columnwidth - 6\tabcolsep) * \real{0.2500}}
  >{\raggedright\arraybackslash}p{(\columnwidth - 6\tabcolsep) * \real{0.2500}}
  >{\raggedright\arraybackslash}p{(\columnwidth - 6\tabcolsep) * \real{0.2500}}
  >{\raggedright\arraybackslash}p{(\columnwidth - 6\tabcolsep) * \real{0.2500}}@{}}
\toprule\noalign{}
\begin{minipage}[b]{\linewidth}\raggedright
\end{minipage} & \begin{minipage}[b]{\linewidth}\raggedright
RMSE at \emph{t} = 5
\end{minipage} & \begin{minipage}[b]{\linewidth}\raggedright
reported posterior SD
\end{minipage} & \begin{minipage}[b]{\linewidth}\raggedright
recovered gain (true 0.551)
\end{minipage} \\
\midrule\noalign{}
\endhead
\bottomrule\noalign{}
\endlastfoot
fresh prior each occasion & 0.1215 & 0.1001 & 0.407 \\
\textbf{posterior carried forward} & \textbf{0.2295} & 0.0571 & 0.281 \\
carried forward, variance inflated & 0.1063 & 0.0795 & 0.410 \\
\end{longtable}

\emph{(n = 1200, six occasions, five items each, items anchored;
tempering exponent d = 0.6. These three conditions are from one run and
are comparable with each other; the runs reported in \S{}5.4 use a
different sample and are not directly comparable to this table.)}

Carrying the posterior forward is worse than having no memory at all,
and it is worse in the way that matters least forgivingly: the reported
uncertainty falls while the actual error rises, so the estimate is
\textbf{wrong by a factor of four relative to what it claims}.

The reason is not subtle once stated. Bayesian updating is a theorem
about accumulating evidence on a \emph{fixed} quantity. A learner's
\(\theta\) is not fixed; that is the point of measuring them repeatedly.
Carrying the posterior forward asserts ``this person is where they were
last time'' as prior information, and for a learner in the middle of
instruction that assertion is false. The estimator then spends the new
evidence overcoming its own prior.

Inflating the variance before carrying forward --- tempering the
posterior by raising it to a power \emph{d} \textless{} 1, a discount
device in the sense of West and Harrison (1997) --- restores the
ordering (third row). But tempering only widens the distribution; it
does not move it. If we know learners are improving, discarding that
knowledge is itself a loss.

What is needed is the \textbf{prediction step} of a state-space model
(West \& Harrison, 1997): a transition that moves the distribution as
well as widening it.

\begin{center}\rule{0.5\linewidth}{0.5pt}\end{center}

\hypertarget{growth-as-a-second-level}{%
\subsection{5.2 Growth as a second
level}\label{growth-as-a-second-level}}

We therefore write a two-level model. The measurement level is the link
function of Section 2 with item parameters anchored; the growth level
describes how \(\theta\) moves:

\[y_{ijt} \sim \text{Bernoulli}\big(P(\theta_{it};\alpha_j,\beta_j,\gamma_j)\big),
\qquad \theta_{it} = g(t;\,\boldsymbol{\psi}_i).\]

Measuring change within item response theory has a literature of its
own, from longitudinal Rasch formulations (Andersen, 1985) and
multidimensional change models (Embretson, 1991) to state-space item
response models estimated by forward filtering (Wang, Berger, \&
Burdick, 2013), with knowledge tracing and ability tracking as the
adaptive-instruction counterparts (Corbett \& Anderson, 1995; Brinkhuis
\& Maris, 2019). Our aim is narrower: a prediction step that reuses the
family's own link function.

Estimation proceeds occasion by occasion. At occasion \emph{t} the
current prior is combined with the responses to give a posterior; the
sequence of posterior means and standard deviations up to \emph{t} is
then used to fit \emph{g}, and the fitted curve evaluated at \emph{t} +
1, with an appropriate predictive variance, becomes the next prior. The
predictive distribution is placed on the grid as a logit-normal, so the
machinery of Section 3 is unchanged.

The growth function \emph{g} is deliberately a plug-in. Nothing in the
construction requires a particular shape, and the shape that is right
will differ between a spaced-practice intervention and a semester of
instruction. We compare three:

\begin{itemize}
\tightlist
\item
  \textbf{random walk} --- \(\theta\)\(_{t+1}\) = \(\theta\)\(_{t}\)
  plus process noise; the tempered carry-forward of \S{}5.1 in another
  guise
\item
  \textbf{logit-linear} --- logit \(\theta\) linear in \emph{t}, the
  natural first choice
\item
  \textbf{four-parameter HCTM} --- the subject of \S{}5.3
\end{itemize}

A caution that turns out to be decisive is recorded here and
demonstrated in \S{}5.5: the number of free growth parameters must lag
the length of the history.

\begin{center}\rule{0.5\linewidth}{0.5pt}\end{center}

\hypertarget{the-link-function-as-a-growth-curve}{%
\subsection{5.3 The link function as a growth
curve}\label{the-link-function-as-a-growth-curve}}

The half-truncated member derived in \S{}2.3 was constructed for a
variable with an absolute origin and no ceiling. Time is such a
variable. Using it as the growth curve, with the ceiling freed to a
learner-specific asymptote \(\delta\),

\[\theta(t) \;=\; \gamma + (\delta-\gamma)\,P_{\infty}(t;\alpha,\beta),
\qquad t \in [0,\infty)
\tag{5.1}\]

gives four parameters, each with a reading a teacher would recognize:

\begin{longtable}[]{@{}ll@{}}
\toprule\noalign{}
parameter & meaning \\
\midrule\noalign{}
\endhead
\bottomrule\noalign{}
\endlastfoot
\(\gamma\) & level at entry \\
\(\delta\) & level the learner converges to \\
\(\alpha\) & rate of learning \\
\(\beta\) & when the growth occurs \\
\end{longtable}

Two features are worth noting. First, \textbf{no horizon constant is
required.} A bounded growth curve would need time rescaled to {[}0, 1{]}
by some arbitrary \emph{T}, and the model would be undefined past it;
(5.1) takes \emph{t} in its natural units. Second, \textbf{\(\delta\) is
not constrained to exceed \(\gamma\)}, so a decreasing curve ---
forgetting --- is expressed by the same four parameters rather than
requiring a separate model.

The measurement level and the growth level thus use one function
evaluated at different anchors: \emph{L} = 1 for the trait, \emph{L}
\(\to\) \(\infty\) for time. They share the quadrature grid, and in an
implementation they share the code.

\begin{center}\rule{0.5\linewidth}{0.5pt}\end{center}

\hypertarget{what-the-prediction-step-buys}{%
\subsection{5.4 What the prediction step
buys}\label{what-the-prediction-step-buys}}

With items anchored, 600 simulated learners were followed over six
occasions of five items each --- a formative-assessment scale, not a
testing scale. All four conditions below come from the same generated
data, so the columns are directly comparable.

\textbf{Accuracy} (RMSE of \(\hat{\theta}\) against truth)

\begin{longtable}[]{@{}lllll@{}}
\toprule\noalign{}
occasion & fresh & logit-linear & CTM growth (\emph{L} = 1) & HCTM
growth \\
\midrule\noalign{}
\endhead
\bottomrule\noalign{}
\endlastfoot
0 & 0.1155 & 0.1155 & 0.1155 & 0.1155 \\
1 & 0.1168 & 0.1197 & 0.1121 & \textbf{0.1059} \\
2 & 0.1128 & 0.1116 & 0.1128 & \textbf{0.0897} \\
3 & 0.1241 & 0.0937 & 0.1079 & \textbf{0.0928} \\
4 & 0.1165 & \textbf{0.0743} & 0.0852 & 0.0872 \\
5 & 0.1444 & \textbf{0.0677} & 0.0838 & 0.0801 \\
\end{longtable}

\textbf{Calibration} (reported posterior SD \(\div\) actual RMSE; 1.00
is honest, below 1 is overconfident)

\begin{longtable}[]{@{}lllll@{}}
\toprule\noalign{}
occasion & fresh & logit-linear & CTM growth & HCTM growth \\
\midrule\noalign{}
\endhead
\bottomrule\noalign{}
\endlastfoot
1 & 1.068 & 0.861 & 1.059 & 1.108 \\
2 & 1.020 & 0.837 & 0.824 & \textbf{1.021} \\
3 & 0.990 & 0.751 & 0.922 & \textbf{1.036} \\
4 & 0.908 & 0.733 & 1.056 & 0.959 \\
5 & 0.812 & 0.659 & 0.986 & \textbf{1.003} \\
\end{longtable}

Two quantities are being reported and they do not order the methods the
same way. The logit-linear prior ends with the lowest error; the HCTM
prior is the only one whose reported uncertainty tracks its actual error
throughout, staying within 0.96--1.11 of unity at every occasion while
the logit-linear figure falls to 0.66 --- a claimed precision a third
better than the one delivered. The fresh prior, having no memory,
degrades in both respects as the cohort moves away from Beta(2, 2).

For an adaptive system the second table is the relevant one --- of the
two requirements set out in \S{}1.3, it is the second, honest
uncertainty, that this table scores. The posterior standard deviation is
not a decoration: it decides when to stop testing, whether a mastery
claim is warranted, and which item to administer next. An estimator
slightly more accurate on average while overstating its precision by a
third is the worse instrument.

Recovery of individual growth (true mean gain 0.553 over the six
occasions) is comparable among the three growth priors and clearly
better than no memory: gain RMSE 0.217 (fresh), 0.153 (logit-linear),
0.157 (CTM), 0.154 (HCTM), with correlations against true gain of 0.687,
0.692, 0.717 and 0.704.

The accuracy of the growth priors also improves as history accumulates,
which no memoryless scheme can do. In a separate run with a longer
sequence the logit-linear prior falls monotonically from 0.1438 to
0.0858 between the first and sixth occasions, while fresh and
random-walk conditions oscillate around 0.13--0.17 indefinitely.
Individual slopes become estimable from about the third occasion; before
that they are shrunk toward the cohort median, an empirical-Bayes device
that carries the first two.

Storage is negligible. Summarizing the carried-forward posterior by a
Beta distribution matched on its first two moments --- \textbf{two
numbers per learner} --- costs at most 0.061 in estimated \(\theta\)
relative to carrying all 41 grid weights, and 0.015 on average.

\begin{center}\rule{0.5\linewidth}{0.5pt}\end{center}

\hypertarget{versatility-and-its-boundary}{%
\subsection{5.5 Versatility, and its
boundary}\label{versatility-and-its-boundary}}

A four-parameter growth curve invites the objection that it will fit
anything. The objection is worth taking seriously in both directions: a
form that fits everything discriminates nothing, and a form that fits
too little fails silently when the assumed shape is wrong.

Fitted to noiseless canonical trajectories, the maximum absolute
approximation error is as follows (the fits are overlaid in Figure 4):

\includegraphics[width=1\textwidth,height=\textheight]{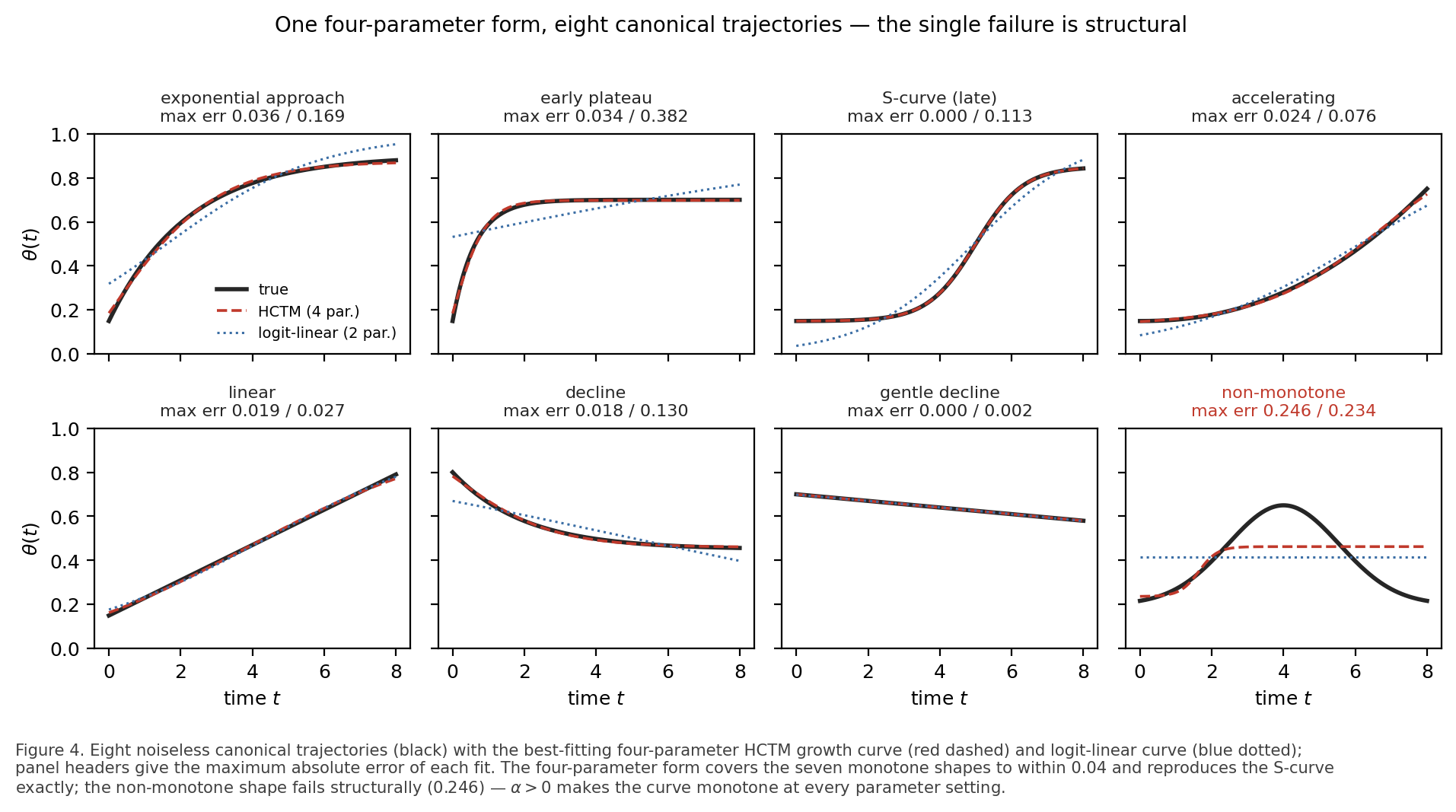}

\begin{longtable}[]{@{}lll@{}}
\toprule\noalign{}
true trajectory & HCTM (4 par.) & logit-linear (2 par.) \\
\midrule\noalign{}
\endhead
\bottomrule\noalign{}
\endlastfoot
exponential approach (decelerating) & 0.036 & 0.169 \\
early plateau & \textbf{0.034} & 0.382 \\
S-curve (late inflection) & \textbf{0.000} & 0.113 \\
accelerating (convex) & 0.024 & 0.076 \\
linear & 0.019 & 0.027 \\
decline (forgetting) & \textbf{0.018} & 0.130 \\
gentle decline & 0.000 & 0.002 \\
\textbf{non-monotone (rise then fall)} & \textbf{0.246} & 0.234 \\
\end{longtable}

\textbf{One four-parameter form covers seven shapes to within 0.04.} The
exact fit to the S-curve is not a coincidence: the curve \emph{is} a
logistic segment, and this is the family's own shape. Decelerating and
accelerating growth are separated by \(\beta\), plateaus by \(\delta\),
and decline by \(\delta\) \textless{} \(\gamma\).

\textbf{The one failure is structural.} \(\alpha\) \textgreater{} 0
makes the curve monotone, so a rise followed by a fall cannot be
represented at any parameter setting. This is not a limitation that more
data would remove.

Under estimation the picture is the same, with one addition:

\begin{longtable}[]{@{}
  >{\raggedright\arraybackslash}p{(\columnwidth - 6\tabcolsep) * \real{0.2500}}
  >{\raggedright\arraybackslash}p{(\columnwidth - 6\tabcolsep) * \real{0.2500}}
  >{\raggedright\arraybackslash}p{(\columnwidth - 6\tabcolsep) * \real{0.2500}}
  >{\raggedright\arraybackslash}p{(\columnwidth - 6\tabcolsep) * \real{0.2500}}@{}}
\toprule\noalign{}
\begin{minipage}[b]{\linewidth}\raggedright
true trajectory
\end{minipage} & \begin{minipage}[b]{\linewidth}\raggedright
fresh
\end{minipage} & \begin{minipage}[b]{\linewidth}\raggedright
logit-linear
\end{minipage} & \begin{minipage}[b]{\linewidth}\raggedright
HCTM
\end{minipage} \\
\midrule\noalign{}
\endhead
\bottomrule\noalign{}
\endlastfoot
exponential approach & 0.143 (0.73) & \textbf{0.059} (0.67) & 0.079
(0.86) \\
plateau then late growth & \textbf{0.127} (0.92) & 0.263 (0.28) & 0.143
(0.57) \\
early plateau & 0.114 (1.05) & 0.178 (0.28) & \textbf{0.091} (1.00) \\
decline & 0.110 (1.12) & 0.109 (0.65) & \textbf{0.106} (0.79) \\
non-monotone & \textbf{0.128} (0.94) & 0.188 (0.34) & 0.142 (0.62) \\
\end{longtable}

\emph{(late-occasion RMSE, with SD/RMSE calibration in parentheses)}

In the two rows where the sequential machinery fails --- non-monotone
trajectories, and plateaus followed by late growth --- \textbf{carrying
no memory is better than carrying a fitted curve.} The reasons differ.
For non-monotone trajectories the failure is structural, as above. For
the late-inflection case it is informational: during a plateau there is
nothing in the history that anticipates the coming rise, and no growth
model of any form could extract it.

The practical asymmetry is the last row of the table read against the
first. \textbf{When the assumed shape is wrong, the flexible form
degrades by 12\% relative to having no memory, while the rigid one
degrades by 100\%.} Four parameters look like the riskier choice and are
in fact the safer one: the risk they carry is overfitting, which \S{}5.6
controls, whereas the risk of two parameters is misspecification, which
nothing controls.

\begin{center}\rule{0.5\linewidth}{0.5pt}\end{center}

\hypertarget{the-condition-of-use-degrees-of-freedom}{%
\subsection{5.6 The condition of use: degrees of
freedom}\label{the-condition-of-use-degrees-of-freedom}}

The sequential scheme has one failure mode severe enough to state as a
rule.

If the number of free growth parameters is released at the same rate as
the history accumulates --- two parameters at two occasions, three at
three, four at four --- the fitted curve interpolates the history
exactly. The residual variance is then zero, the predictive variance
collapses, and the next prior is a spike. That spike dominates the next
occasion's likelihood, producing an estimate that is confidently wrong;
the error compounds.

We reproduced this failure, and it is severe: RMSE 0.21 with the
reported standard deviation seven times smaller than the actual error
--- worse than every other condition tested, including no memory at all.

The remedy is to keep at least one degree of freedom in reserve. With
\emph{H} the number of occasions in the history and \emph{k} the number
of free growth parameters,

\begin{quote}
\textbf{release parameters so that \emph{k} \(\le\) \emph{H} \(-\) 1},
and floor the residual variance at the measurement error implied by the
posterior standard deviations.
\end{quote}

Under this schedule --- \(\gamma\) alone at \emph{H} \(\le\) 2, adding
\(\delta\) at \emph{H} = 3, \(\alpha\) at \emph{H} = 4, \(\beta\) at
\emph{H} \(\ge\) 5 --- the calibration figures of \S{}5.4 are obtained.
Without it the same model produces the worst results in this paper.

\begin{center}\rule{0.5\linewidth}{0.5pt}\end{center}

\hypertarget{summary}{%
\subsection{5.7 Summary}\label{summary}}

Bayesian updating alone is not enough for a learner who is changing, and
applied naively it is actively harmful. Adding a prediction step in
which a growth curve is fitted to the learner's own history restores the
benefit and, with a sufficiently flexible curve, delivers uncertainty
that matches the error actually made. The half-truncated member of the
family serves as that curve without requiring a horizon constant, and
covers accelerating, decelerating, plateauing and declining trajectories
in one form --- but not non-monotone ones, and not without a
degrees-of-freedom discipline.

\begin{center}\rule{0.5\linewidth}{0.5pt}\end{center}

\hypertarget{an-empirical-illustration-lsat-section-6}{%
\section{6. An Empirical Illustration: LSAT Section
6}\label{an-empirical-illustration-lsat-section-6}}

\begin{center}\rule{0.5\linewidth}{0.5pt}\end{center}

\hypertarget{the-data}{%
\subsection{6.1 The data}\label{the-data}}

We use the responses of 1000 examinees to the five dichotomously scored
items of Section 6 of the Law School Admission Test, first reported by
Bock and Lieberman (1970) and analysed by Choi (2022) as the empirical
illustration of the three-parameter Cognitive Trait Model. The data are
distributed as 32 response patterns with frequencies. All items are
five-option multiple choice, which is what makes a lower asymptote
meaningful.

The response matrix was reconstructed from the pattern table and checked
against the published data before use: the 32 pattern frequencies agree
exactly, the item proportions correct are 0.924, 0.709, 0.553, 0.763 and
0.870, and the frequencies sum to 1000. Two features of the data will
matter throughout: \textbf{the test is easy} (298 examinees, 29.8\%,
answer all five items correctly) and \textbf{it is short}.

\begin{center}\rule{0.5\linewidth}{0.5pt}\end{center}

\hypertarget{reproducing-the-published-item-parameters}{%
\subsection{6.2 Reproducing the published item
parameters}\label{reproducing-the-published-item-parameters}}

Fitting the three-parameter model by Bayes modal EM with the priors of
the source article --- \(\theta\) \(\sim\) Beta(2, 2), \(\beta\)
\(\sim\) Beta(2, 2), \(\gamma\) \(\sim\) Beta(7, 25) --- and a
log-normal prior on \(\alpha\) gives:

\begin{longtable}[]{@{}
  >{\raggedright\arraybackslash}p{(\columnwidth - 12\tabcolsep) * \real{0.1429}}
  >{\raggedright\arraybackslash}p{(\columnwidth - 12\tabcolsep) * \real{0.1429}}
  >{\raggedright\arraybackslash}p{(\columnwidth - 12\tabcolsep) * \real{0.1429}}
  >{\raggedright\arraybackslash}p{(\columnwidth - 12\tabcolsep) * \real{0.1429}}
  >{\raggedright\arraybackslash}p{(\columnwidth - 12\tabcolsep) * \real{0.1429}}
  >{\raggedright\arraybackslash}p{(\columnwidth - 12\tabcolsep) * \real{0.1429}}
  >{\raggedright\arraybackslash}p{(\columnwidth - 12\tabcolsep) * \real{0.1429}}@{}}
\toprule\noalign{}
\begin{minipage}[b]{\linewidth}\raggedright
item
\end{minipage} & \begin{minipage}[b]{\linewidth}\raggedright
\(\alpha\) published
\end{minipage} & \begin{minipage}[b]{\linewidth}\raggedright
\(\alpha\) here
\end{minipage} & \begin{minipage}[b]{\linewidth}\raggedright
\(\beta\) published
\end{minipage} & \begin{minipage}[b]{\linewidth}\raggedright
\(\beta\) here
\end{minipage} & \begin{minipage}[b]{\linewidth}\raggedright
\(\gamma\) published
\end{minipage} & \begin{minipage}[b]{\linewidth}\raggedright
\(\gamma\) here
\end{minipage} \\
\midrule\noalign{}
\endhead
\bottomrule\noalign{}
\endlastfoot
1 & 8.86 & 7.97 & 0.08 & 0.05 & 0.33 & 0.32 \\
2 & 1.25 & 0.53 & 0.43 & 0.46 & 0.38 & 0.40 \\
3 & 2.61 & 2.46 & 0.67 & 0.78 & 0.20 & 0.21 \\
4 & 1.98 & 2.85 & 0.35 & 0.18 & 0.45 & 0.38 \\
5 & 5.21 & 5.14 & 0.13 & 0.08 & 0.44 & 0.40 \\
\end{longtable}

The lower asymptotes agree closely for all five items. The slopes and
locations agree for items 1, 3 and 5 and diverge for items 2 and 4. Some
difference is expected --- the estimators differ, and we use a different
prior on \(\alpha\) --- but the size of the discrepancy for item 2
(\(\alpha\) of 1.25 against 0.53) is better read as evidence about the
data than about either estimator. \textbf{Fifteen item parameters are
being estimated from five binary items}, and \S{}6.4 shows directly that
the likelihood is nearly flat in several of them.

The estimates are otherwise well behaved: the EM converges in 94
iterations, and the two-parameter model fitted to the same data gives a
log-likelihood of \(- 2520.24\) against \(- 2477.29\) for the
three-parameter model.

\begin{center}\rule{0.5\linewidth}{0.5pt}\end{center}

\hypertarget{the-published-correlation-table}{%
\subsection{6.3 The published correlation
table}\label{the-published-correlation-table}}

The source article reports Spearman correlations among \(\theta\)
estimates from four models, observing that in the simulated data the 2PL
and 2P CTM orderings were identical whereas in the LSAT data they are
not, and attributing the difference to test length. Our reproduction
agrees in direction:

\begin{longtable}[]{@{}lll@{}}
\toprule\noalign{}
model pair & published \(\rho\) & here \\
\midrule\noalign{}
\endhead
\bottomrule\noalign{}
\endlastfoot
2PL -- 2P CTM & 0.869 & 0.920 \\
2PL -- 3PL & 0.948 & 0.996 \\
2PL -- 3P CTM & 0.922 & 0.944 \\
2P CTM -- 3PL & 0.896 & 0.926 \\
2P CTM -- 3P CTM & 0.918 & 0.974 \\
3PL -- 3P CTM & 0.957 & 0.950 \\
\end{longtable}

All correlations are below 1, as reported. The mechanism, however, is
more specific than test length as such. With five binary items there are
32 possible response patterns and 30 occur; a thousand examinees
therefore receive at most 30 distinct estimates, and the ranking is
dominated by ties. Among the ten patterns with a total score of 3, the
2PL model ranks \texttt{10101} highest while the 2P CTM model ranks
\texttt{10011} highest --- a disagreement over a handful of patterns,
each carrying many examinees, is enough to move a rank correlation
appreciably. \textbf{It is not the number of items but the coarseness of
the resulting partition that produces the effect}, which is why the
simulated data with twenty items showed none.

This is also an instance of the caution in \S{}4.3. A correlation of
0.92 between two models is not evidence that they measure nearly the
same thing when only 30 values are available to correlate.

\begin{center}\rule{0.5\linewidth}{0.5pt}\end{center}

\hypertarget{what-five-items-can-and-cannot-identify}{%
\subsection{6.4 What five items can and cannot
identify}\label{what-five-items-can-and-cannot-identify}}

The three-parameter model is applied here because the items are multiple
choice. The question is whether the data support it.

We refitted the model under four priors for \(\gamma\), from the
informative Beta(7, 25) used in the source article (mode 0.2, matching
one over the number of options) to a uniform Beta(1, 1). If the data
identify \(\gamma\), the estimates should be stable across these.

\begin{longtable}[]{@{}
  >{\raggedright\arraybackslash}p{(\columnwidth - 10\tabcolsep) * \real{0.1667}}
  >{\raggedright\arraybackslash}p{(\columnwidth - 10\tabcolsep) * \real{0.1667}}
  >{\raggedright\arraybackslash}p{(\columnwidth - 10\tabcolsep) * \real{0.1667}}
  >{\raggedright\arraybackslash}p{(\columnwidth - 10\tabcolsep) * \real{0.1667}}
  >{\raggedright\arraybackslash}p{(\columnwidth - 10\tabcolsep) * \real{0.1667}}
  >{\raggedright\arraybackslash}p{(\columnwidth - 10\tabcolsep) * \real{0.1667}}@{}}
\toprule\noalign{}
\begin{minipage}[b]{\linewidth}\raggedright
\(\gamma\) prior (mode)
\end{minipage} & \begin{minipage}[b]{\linewidth}\raggedright
item 1
\end{minipage} & \begin{minipage}[b]{\linewidth}\raggedright
item 2
\end{minipage} & \begin{minipage}[b]{\linewidth}\raggedright
item 3
\end{minipage} & \begin{minipage}[b]{\linewidth}\raggedright
item 4
\end{minipage} & \begin{minipage}[b]{\linewidth}\raggedright
item 5
\end{minipage} \\
\midrule\noalign{}
\endhead
\bottomrule\noalign{}
\endlastfoot
Beta(7, 25) --- 0.20 & 0.324 & 0.396 & 0.215 & 0.381 & 0.402 \\
Beta(4, 16) --- 0.17 & 0.357 & 0.402 & 0.173 & 0.436 & 0.506 \\
Beta(2, 8) --- 0.13 & 0.532 & 0.409 & 0.109 & 0.515 & 0.651 \\
Beta(1, 1) --- uniform & \textbf{0.820} & 0.416 & \textbf{0.104} & 0.525
& \textbf{0.739} \\
\emph{proportion correct} & \emph{0.924} & \emph{0.709} & \emph{0.553} &
\emph{0.763} & \emph{0.870} \\
\end{longtable}

Items 2 and 3 are stable; items 1 and 5 are not. The pattern is ordered
by difficulty, and the explanation is direct. \textbf{\(\gamma\) is the
probability of a correct response at \(\theta\) = 0, so the data speak
to it only through examinees near the bottom of the scale who answer the
item incorrectly.} Item 1 is answered correctly by 92.4\% of the sample;
the failures that would inform \(\gamma\)\(_{1}\) are too few, and the
estimate follows the prior. Under a uniform prior \(\gamma\)\(_{1}\) =
0.82 --- the assertion that an examinee at the ignorance level has an
82\% chance of answering a five-option item correctly, which is not
credible on any reading.

A simulation isolates the threshold. Generating from a known \(\gamma\)
= 0.2 and comparing estimates under an informative and a uniform prior,
the mean absolute difference is:

\begin{longtable}[]{@{}ll@{}}
\toprule\noalign{}
item proportion correct & prior sensitivity \\
\midrule\noalign{}
\endhead
\bottomrule\noalign{}
\endlastfoot
\(\ge\) 0.90 & \textbf{0.214} \\
0.80 -- 0.90 & 0.035 \\
0.60 -- 0.80 & 0.022 \\
\textless{} 0.60 & 0.002 \\
\end{longtable}

The deterioration is not gradual but concentrated above roughly 0.90,
and LSAT items 1 (0.924) and 5 (0.870) sit on either side of that
boundary. At a proportion correct of 0.974 the estimate under a uniform
prior is 0.611 against a true 0.2.

\textbf{Lengthening the test does not help.} Holding the target items
fixed and adding filler items to raise \emph{J} from 4 to 40 leaves the
prior sensitivity of the target items unchanged (0.231 to 0.232), even
though the posterior standard deviation of \(\theta\) for low-ability
examinees falls by 61\% over the same range. Increasing the sample does
help: with \emph{J} fixed, raising \emph{n} from 1000 to 64,000 reduces
the sensitivity from 0.310 to 0.033. The binding constraint is the
\textbf{number of observed failures near \(\theta\) = 0}, which grows
with the number of examinees and not with the number of items --- an
item's lower asymptote is its own parameter, and other items carry no
information about it.

Two consequences for practice. Items with a proportion correct above
about 0.90 should not have \(\gamma\) estimated from the data; fix it at
one over the number of options, or use the two-parameter model. And
\(\gamma\) estimates should never be reported without the prior that
produced them, with a sensitivity analysis alongside.

\textbf{This difficulty is not peculiar to the bounded model.} The
corresponding parameter of the three-parameter logistic model is at
least as badly determined --- a difficulty documented since its
introduction (Lord, 1980) --- and in a matched comparison the mean prior
sensitivity is 0.140 for the 3PL lower asymptote against 0.103 for the
bounded model's \(\gamma\), and the gap is widest for items of middling
difficulty (0.132 against 0.009). The reason is structural: the 3PL
asymptote is the limit as \(\theta\) \(\to\) \(- \infty\), a point no
examinee occupies, so it is always an extrapolation, whereas \(\gamma\)
is the value at an attainable point of the scale. The bounded model
inherits a known difficulty in a slightly milder form.

\begin{center}\rule{0.5\linewidth}{0.5pt}\end{center}

\hypertarget{scoring-a-short-easy-test}{%
\subsection{6.5 Scoring a short, easy
test}\label{scoring-a-short-easy-test}}

The features noted in \S{}6.1 come together in the person estimates. Of
the 1000 examinees, 323 (32.3\%) receive a maximum likelihood estimate
at a boundary --- a rate far above the 14.5\% our simulations predicted
for five items, because the simulated tests were not this easy.

For the 298 examinees who answered every item correctly:

\begin{longtable}[]{@{}lll@{}}
\toprule\noalign{}
estimator & estimate & interpretation offered \\
\midrule\noalign{}
\endhead
\bottomrule\noalign{}
\endlastfoot
EAP & 0.666 (posterior SD 0.179) & 67\% of mastery \\
MAP & 0.749 & 75\% of mastery \\
MLE & 1.000 & \textbf{100\% of mastery} \\
\end{longtable}

The three readings differ by a third of the scale. In the twenty-item
simulation of \S{}3.3 the same three estimators agreed to within 0.07;
here they do not, and the posterior standard deviation of 0.179
indicates that no point estimate should be reported alone for these
examinees.

The full picture by total score:

\begin{longtable}[]{@{}llllll@{}}
\toprule\noalign{}
total score & \emph{n} & EAP & posterior SD & MAP & MLE \\
\midrule\noalign{}
\endhead
\bottomrule\noalign{}
\endlastfoot
0 & 3 & 0.139 & 0.086 & 0.087 & 0.000 \\
1 & 20 & 0.204 & 0.114 & 0.149 & 0.022 \\
2 & 85 & 0.284 & 0.141 & 0.233 & 0.118 \\
3 & 237 & 0.391 & 0.166 & 0.357 & 0.289 \\
4 & 357 & 0.515 & 0.181 & 0.520 & 0.535 \\
5 & 298 & 0.666 & 0.179 & 0.749 & \textbf{1.000} \\
\end{longtable}

The posterior standard deviation rises with the total score up to 0.18
and does not fall at the top, which is the behaviour \S{}3.3 predicts
and the Fisher standard error does not: information diverges as
\(\theta\) \(\to\) 1, so an information-based standard error would
report near-perfect precision for exactly the examinees about whom least
is known.

\textbf{This is the paper's central practical caution in empirical
form.} The proportion-of-mastery reading is the model's principal
attraction, and on a short easy test it is not determined to better than
a third of the scale by the data alone. The setting is not exotic: five
to ten items administered to a cohort that has largely mastered the
material is exactly the formative assessment for which the model was
designed.

The recommendations of \S{}3.3 follow directly. Report MAP with the EAP
posterior standard deviation. Do not report the maximum likelihood
estimate as a trait level; flag its boundary solutions instead. And do
not attach the phrase ``100\% of mastery'' to an examinee whose data are
consistent with anything above about 0.5.

\begin{center}\rule{0.5\linewidth}{0.5pt}\end{center}

\hypertarget{summary-1}{%
\subsection{6.6 Summary}\label{summary-1}}

The LSAT analysis reproduces the published item estimates for the
parameters that the data determine, and shows which those are. The lower
asymptotes of the two easiest items are set by the prior rather than by
the data; the three point estimators disagree by a third of the scale
for the 30\% of examinees with perfect scores; and the model-to-model
rank correlations reported in the source article are shaped by the
coarseness of a 30-value partition rather than by measurement
differences. None of this impugns the model. It locates the conditions
under which its interpretive claim can be made good --- enough items
that perfect scores are rare, enough low-performing examinees that lower
asymptotes are identified, and an uncertainty reported alongside every
point estimate.

\begin{center}\rule{0.5\linewidth}{0.5pt}\end{center}

\hypertarget{practical-guidance}{%
\section{7. Practical Guidance}\label{practical-guidance}}

Twelve recommendations follow from the results above. Each is stated
with the finding that supports it, so that a reader who disagrees with
the reasoning can locate the evidence.

All twelve are implemented as the defaults of an accompanying
open-source package (Choi, 2026a), so that following them requires no
additional decision by the user: scoring returns the MAP estimate
together with the posterior standard deviation and flags boundary
solutions, and the degrees-of-freedom rule of recommendation 9 is
enforced inside the growth fit rather than offered as an option. The
same estimator also runs in a browser, where the warnings of
recommendations 3 to 6 are raised on the reader's own data with nothing
installed and nothing uploaded (Choi, 2026b); the calibration of Section
6 can be reproduced from its opening screen, including the item whose
lower asymptote moves from 0.32 to 0.82 when the prior is flattened.

\textbf{1. Do not use the maximum likelihood estimate as a reported
trait level.} Three independent lines of evidence converge here. It
produces boundary solutions in 4.8\% of examinees at \(J = 20\) in the
two-parameter model, 5.7\% in the three-parameter model at the same
length, 14.5\% and 23.5\% respectively at \(J = 5\), and \textbf{32.3\%
in the LSAT data} (\S{}3.3, \S{}6.5). Unlike the \(\pm\)\(\infty\) of
item response theory, the boundary value here is \(\hat{\theta}\) = 1,
which reads as ``100\% of mastery'' and does not announce itself as a
failure. In the three-parameter model the failure is worse: of 103
boundary solutions at \(J = 20\), only one belonged to an examinee who
answered every item incorrectly, so the extreme-pattern signal that
permits screening in the two-parameter model is absent (\S{}3.3). Report
\textbf{MAP as the point estimate and the EAP posterior standard
deviation as the uncertainty}; they share a grid and cost nothing extra.
Flag boundary solutions rather than reporting them.

\textbf{2. Use the same rule for interim estimates in adaptive testing.}
With EAP as the interim estimate, information-based item selection
performs identically whether or not the information function is
truncated. With MLE it is actively harmful: examinees driven to the
boundary receive zero truncated information at every item, selection
becomes effectively random, and the RMSE among high performers more than
doubles (\S{}3.5). If a selection rule is being chosen at the same time,
expected posterior variance minimization is slightly more accurate than
maximum information (0.0528 against 0.0569), immune to the boundary
divergence by construction, and costs 22.9 \(\mu\)s per item against
12.1 \(\mu\)s --- still 40,000 selections per second.

\textbf{3. Do not estimate the lower asymptote for items with a
proportion correct above about 0.90.} \(\gamma\) is the probability of
success at \(\theta\) = 0, so the data inform it only through
low-ability examinees who answer incorrectly. Above roughly 0.90 those
observations become too few and the estimate follows the prior: changing
the prior from Beta(7, 25) to uniform moves \(\hat{\gamma}\) by 0.214 on
average for such items, against 0.002 for items below 0.60 (\S{}6.4).
Fix \(\gamma\) at one over the number of options, or use the
two-parameter model. The cost is small --- fixing \(\gamma\) at its true
value rather than estimating it changes person scoring by 0.0006 in RMSE
(\S{}3.3).

\textbf{4. Never report \(\hat{\gamma}\) without the prior that produced
it}, and include a sensitivity analysis in routine output. In the LSAT
data the estimate for item 1 ranges from 0.32 to 0.82 across reasonable
priors (\S{}6.4).

\textbf{5. Lengthening the test does not identify \(\gamma\); enlarging
the sample does.} Holding target items fixed while raising \(J\) from 4
to 40 leaves prior sensitivity unchanged, whereas raising \(n\) from
1,000 to 64,000 reduces it from 0.310 to 0.033 (\S{}6.4). An item's
lower asymptote is its own parameter; other items carry no information
about it.

\textbf{6. Report an uncertainty with every point estimate on short
tests.} For the 298 LSAT examinees with perfect scores the posterior
standard deviation is 0.179 and the three estimators disagree by a third
of the scale (\S{}6.5). The Fisher standard error will not serve:
information diverges as \(\theta\) \(\to\) 1, so it reports near-perfect
precision for exactly the examinees about whom least is known (\S{}3.5).

\textbf{7. Forty-one quadrature nodes suffice.} Estimates agree with a
61-node grid to 4 \(\times\) 10\(^{-}\)\(^{1}\)\(^{5}\); twenty-one
nodes agree to 1 \(\times\) 10\(^{-}\)\(^{8}\) and are adequate in
practice. Eleven are not (\S{}3.1).

\textbf{8. Design items with \(\alpha\) roughly in 10--20 if parameter
recovery matters.} Relative recovery error is U-shaped: 51.2\% at
\(\alpha\) = 2, 14.4\% at 5, 9.8\% at 10, 11.0\% at 20, 19.8\% at 40
(mean over 20 replications). Flat curves carry no slope information and
steep ones saturate (\S{}3.2).

\textbf{9. In sequential updating, keep at least one degree of freedom
in reserve.} With \(H\) occasions of history and \(k\) free growth
parameters, release so that \(k \le H - 1\), and floor the residual
variance at the measurement error implied by the posterior standard
deviations. Violating this produces the worst result in this paper: RMSE
0.21 with a reported standard error seven times smaller than the actual
error (\S{}5.6).

\textbf{10. Turn sequential updating off when trajectories may be
non-monotone.} The growth curve is monotone by construction; forcing it
extrapolates in the wrong direction and does worse than having no memory
(\S{}5.5). Post-intervention forgetting and between-term loss are the
obvious cases.

\textbf{11. Screen items before calibration.} The model admits only
\(\alpha\) \textgreater{} 0, so reverse-keyed and aberrant items cannot
be detected by the fit; remove them beforehand using point-biserial
correlation or a comparable statistic; the caution appears in the source
article as well (Choi, 2022).

\textbf{12. Anchor the item parameters if scores are to be compared
across cohorts.} A fixed examinee's estimate moves by 69\% of the score
spread when the surrounding population changes and the items are
recalibrated (\S{}4.5). This requirement is not peculiar to this family
and is noted in the source article (Choi, 2022), but the
criterion-referenced reading makes it easy to forget: ``62\% of
mastery'' sounds absolute in a way that ``\(\theta\) = 0.31'' does not.

\begin{center}\rule{0.5\linewidth}{0.5pt}\end{center}

\hypertarget{discussion}{%
\section{8. Discussion}\label{discussion}}

\hypertarget{what-was-established}{%
\subsection{8.1 What was established}\label{what-was-established}}

The derivation underlying the Cognitive Trait Model imposes two boundary
conditions. Imposing only the first --- the one that fixes the origin
--- yields a family indexed by the mastery anchor \(L\), of which the
published model is the case \(L = 1\) and the unbounded half-truncated
link is the limit \emph{L} \(\to\) \(\infty\). The members share a
kernel and differ by a normalizing constant, and the whole family admits
a single derivative formula, a single information function, and a single
statement about where that information function diverges.

Because the trait has bounded support, estimation does not require MCMC.
Bayes modal estimation on a fixed Gauss-Legendre grid reproduces the
published estimates to a correlation of 0.9999 while running 35 times
faster than a converged random-walk sampler and 73 times faster than
NUTS, and reduces a single adaptive-testing update to about five
microseconds. This is the finding that moves the model from post hoc
analysis into the response loop, and it follows from the same property
that produces the interpretive claim.

Applied to the time axis, the link becomes a four-parameter growth curve
whose parameters correspond to entry level, asymptote, rate and timing,
and which represents accelerating, decelerating, plateauing and
declining trajectories in a single form. Fitted to a learner's own
history it supplies the prior for the next occasion, giving estimates
that improve as history accumulates and --- with the unbounded member
--- uncertainty that matches the error actually made.

\hypertarget{what-was-not}{%
\subsection{8.2 What was not}\label{what-was-not}}

We looked three times for a measurement advantage over item response
theory and did not find one. This is not a tautology: the family reduces
to a reparameterization of the logistic model only when all items share
a common slope, so the two- and three-parameter members are genuinely
different functions of \(\theta\) and a difference was possible in
principle. Where differences do appear they are small,
condition-dependent, and run in both directions.

Nor is the criterion-referenced reading as secure as the construction
suggests. Under marginal estimation the scale is held by the \(\theta\)
prior rather than by the task domain: replacing the item domain with its
easier or harder half moves the mean estimate by less than 1.5\%, while
moving the population shifts a fixed examinee's estimate by 69\% of the
score spread. The source article anticipates instability from the domain
and does not discuss the population; the exposure runs the other way.

A second qualification of the same reading is independent of that one.
The two boundary conditions pin \(\theta\) and the expected proportion
of the domain together at the endpoints and nowhere else, so a
coordinate of 0.62 is not in general a claim that 62\% of the domain can
be performed; where the domain proportion is what is wanted it should be
computed and reported as such (\S{}4.5).

We take neither result as an objection to the framework. The first
locates the contribution in interpretation and computation rather than
in measurement precision. The second is a call to carry the framework's
own logic further --- to anchor explicitly what the construction names.

\hypertarget{limitations}{%
\subsection{8.3 Limitations}\label{limitations}}

\textbf{The speed comparison is against a specific class of sampler.} We
specified, tuned and diagnosed two Metropolis samplers and one
gradient-based sampler, and report effective samples per second rather
than wall clock alone. A well-designed Gibbs sampler exploiting
conjugacy, or a specialized data-augmentation scheme, might narrow the
gap; we did not test one.

\textbf{The growth results assume monotone trajectories and anchored
items.} Both were examined rather than asserted, and \S{}5.5 reports
what happens when the first fails. Online item calibration was not
attempted.

\textbf{Growth parameters were fitted by weighted least squares} on the
sequence of posterior means, not marginalized over a prior. A fully
Bayesian treatment would propagate growth-parameter uncertainty into the
predictive variance and might improve calibration further. Individual
growth rates were shrunk toward the cohort median rather than estimated
hierarchically.

\textbf{The empirical illustration is a single short test.} Five items
and 1000 examinees demonstrate the failure modes but do not establish
typical behaviour. The 32.3\% boundary rate in particular reflects an
unusually easy test.

\textbf{Replication counts differ across results.} Headline comparisons
use 20 replications; the growth simulations use five, and a few
exploratory results are single-run. We report the replication standard
deviation wherever we have it, and the \(\alpha\) results in \S{}3.2
show why: the sampling variability of RMSE(\(\alpha\)) exceeds the
differences one might otherwise be tempted to interpret.

\hypertarget{directions}{%
\subsection{8.4 Directions}\label{directions}}

The most immediate extension is \textbf{multidimensionality}. The
bounded construction generalizes to a vector of traits, and the
quadrature argument survives in low dimensions, but the fixed-grid
advantage decays exponentially and a different computational strategy
would be needed beyond two or three dimensions.

\textbf{Online calibration} is the missing half of the sequential story.
If the machinery of \S{}5 is applied to item parameters as well as to
persons, the item bank becomes a state to be tracked, which is what a
live adaptive system requires.

The \textbf{degrees-of-freedom rule of \S{}5.6} was arrived at
empirically after reproducing its violation. A principled version ---
releasing growth parameters by a criterion rather than a schedule ---
would be a genuine improvement, and model averaging over growth forms
would carry model uncertainty into the predictive variance.

Finally, the practical case rests on claims about adaptive instruction
supported by simulation only. \textbf{A deployment study}, in which the
microsecond update is actually used to select the next task and the
resulting sequence is compared against a fixed one, is the evidence the
argument ultimately requires.

Characterizing the item-bank conditions under which \(\theta\) coincides
with the expected proportion of the domain, and what follows for how a
proportion-of-mastery score should be reported, is left to future work.

\hypertarget{conclusion}{%
\subsection{8.5 Conclusion}\label{conclusion}}

Item response theory freed the trait from the items and, in doing so,
gave the scale's origin and unit to convention. The Cognitive Trait
Model takes them back and hands them to the task domain, and the
resulting number --- a proportion of mastery --- says something a
practitioner can use. What we have added is, first, that only one of
those two anchors is necessary, so the model belongs to a family rather
than standing alone; second, that the same bounded support which
produces the interpretation also removes the estimation obstacle,
placing the model inside the response loop rather than after it; and
third, that the same link function, turned onto the time axis, describes
how a learner moves along the scale it defines.

The origin is what must be secured. Current practice does not secure it,
and saying so is the first step toward doing better.

\begin{center}\rule{0.5\linewidth}{0.5pt}\end{center}

\hypertarget{appendix.-theta-and-the-expected-proportion-of-the-domain}{%
\section{\texorpdfstring{Appendix. \(\theta\) and the expected
proportion of the
domain}{Appendix. \textbackslash theta and the expected proportion of the domain}}\label{appendix.-theta-and-the-expected-proportion-of-the-domain}}

The two boundary conditions \(P(0) = 0\) and \(P(L) = 1\) hold for every
item, so \(\theta\) and the expected proportion of the domain agree
exactly at the two endpoints and are otherwise unconstrained. The
expected proportion at a given \(\theta\) is the item-average of the
response functions,

\[\mathrm{DS}(\theta) \;=\; \frac{1}{J}\sum_j P_j(\theta),\]

which depends on how item difficulty is distributed across the domain.
Evaluated at the LSAT estimates of \S{}6.2 (the ``here'' columns), and
contrasted with banks whose difficulties are spread uniformly:

\begin{longtable}[]{@{}
  >{\raggedright\arraybackslash}p{(\columnwidth - 14\tabcolsep) * \real{0.1250}}
  >{\raggedright\arraybackslash}p{(\columnwidth - 14\tabcolsep) * \real{0.1250}}
  >{\raggedright\arraybackslash}p{(\columnwidth - 14\tabcolsep) * \real{0.1250}}
  >{\raggedright\arraybackslash}p{(\columnwidth - 14\tabcolsep) * \real{0.1250}}
  >{\raggedright\arraybackslash}p{(\columnwidth - 14\tabcolsep) * \real{0.1250}}
  >{\raggedright\arraybackslash}p{(\columnwidth - 14\tabcolsep) * \real{0.1250}}
  >{\raggedright\arraybackslash}p{(\columnwidth - 14\tabcolsep) * \real{0.1250}}
  >{\raggedright\arraybackslash}p{(\columnwidth - 14\tabcolsep) * \real{0.1250}}@{}}
\toprule\noalign{}
\begin{minipage}[b]{\linewidth}\raggedright
\(\theta\)
\end{minipage} & \begin{minipage}[b]{\linewidth}\raggedright
0.00
\end{minipage} & \begin{minipage}[b]{\linewidth}\raggedright
0.20
\end{minipage} & \begin{minipage}[b]{\linewidth}\raggedright
0.40
\end{minipage} & \begin{minipage}[b]{\linewidth}\raggedright
0.50
\end{minipage} & \begin{minipage}[b]{\linewidth}\raggedright
0.60
\end{minipage} & \begin{minipage}[b]{\linewidth}\raggedright
0.80
\end{minipage} & \begin{minipage}[b]{\linewidth}\raggedright
1.00
\end{minipage} \\
\midrule\noalign{}
\endhead
\bottomrule\noalign{}
\endlastfoot
2P, this study & 0.000 & 0.325 & 0.573 & 0.667 & 0.748 & 0.885 &
1.000 \\
3P, this study & 0.342 & 0.553 & 0.714 & 0.776 & 0.829 & 0.922 &
1.000 \\
\(\beta\) uniform, J = 5 & 0.000 & 0.177 & 0.392 & 0.500 & 0.608 & 0.823
& 1.000 \\
\(\beta\) uniform, J = 20 & 0.000 & 0.179 & 0.392 & 0.500 & 0.608 &
0.821 & 1.000 \\
\end{longtable}

The 2P row is the same estimates with \(\gamma\) set to zero rather than
a separately fitted two-parameter model, so that the rows differ only in
the lower asymptote. The two uniform banks, shown for contrast, use
\(\alpha\) = 8 at difficulties \((j - \tfrac12)/J\); \(\theta\) and
DS(\(\theta\)) coincide there to within 0.02, and the agreement is not
improved by lengthening the test. The published parameters of Choi
(2022) give the same picture (0.653 and 0.773 at \(\theta\) = 0.50 for
the two- and three-parameter forms), so the divergence is a property of
the item bank rather than of either set of estimates. No new estimation
is involved: these are the manuscript's own parameters substituted into
the manuscript's own equations.

At the level of individual estimates the three {[}0, 1{]} scales of
\S{}4.5 can be compared directly. For the LSAT data, with the percentile
rank computed by the midpoint convention
\((\text{count below} + n/2)/N\):

\begin{longtable}[]{@{}
  >{\raggedright\arraybackslash}p{(\columnwidth - 8\tabcolsep) * \real{0.2000}}
  >{\raggedright\arraybackslash}p{(\columnwidth - 8\tabcolsep) * \real{0.2000}}
  >{\raggedright\arraybackslash}p{(\columnwidth - 8\tabcolsep) * \real{0.2000}}
  >{\raggedright\arraybackslash}p{(\columnwidth - 8\tabcolsep) * \real{0.2000}}
  >{\raggedright\arraybackslash}p{(\columnwidth - 8\tabcolsep) * \real{0.2000}}@{}}
\toprule\noalign{}
\begin{minipage}[b]{\linewidth}\raggedright
total score
\end{minipage} & \begin{minipage}[b]{\linewidth}\raggedright
n
\end{minipage} & \begin{minipage}[b]{\linewidth}\raggedright
proportion correct
\end{minipage} & \begin{minipage}[b]{\linewidth}\raggedright
\(\hat{\theta}\) (EAP)
\end{minipage} & \begin{minipage}[b]{\linewidth}\raggedright
percentile rank
\end{minipage} \\
\midrule\noalign{}
\endhead
\bottomrule\noalign{}
\endlastfoot
0 & 3 & 0.000 & 0.139 & 0.002 \\
1 & 20 & 0.200 & 0.204 & 0.013 \\
2 & 85 & 0.400 & 0.284 & 0.066 \\
3 & 237 & 0.600 & 0.391 & 0.227 \\
4 & 357 & 0.800 & 0.515 & 0.523 \\
5 & 298 & 1.000 & 0.666 & 0.851 \\
\end{longtable}

The three columns agree on the ordering and on nothing else. At a total
score of 3 the same examinee is at 0.600 of the items, 0.391 of the
trait scale and 0.227 of the cohort. At the level of estimates the
endpoints do not agree either, because shrinkage moves them; the
statement that the endpoints agree is a statement about the scales, not
about estimates from a short test.

\begin{center}\rule{0.5\linewidth}{0.5pt}\end{center}

\hypertarget{references}{%
\section{References}\label{references}}

Andersen, E. B. (1985). Estimating latent correlations between repeated
testings. \emph{Psychometrika}, 50(1), 3--16. doi:10.1007/BF02294143

Birnbaum, A. (1968). Some latent trait models and their use in inferring
an examinee's ability. In F. M. Lord \& M. R. Novick, \emph{Statistical
theories of mental test scores} (pp.~397--479). Reading, MA:
Addison-Wesley.

Black, P., \& Wiliam, D. (1998). Assessment and classroom learning.
\emph{Assessment in Education: Principles, Policy \& Practice}, 5(1),
7--74. doi:10.1080/0969595980050102

Bock, R. D., \& Aitkin, M. (1981). Marginal maximum likelihood
estimation of item parameters: Application of an EM algorithm.
\emph{Psychometrika}, 46(4), 443--459. doi:10.1007/BF02293801

Bock, R. D., \& Lieberman, M. (1970). Fitting a response model for n
dichotomously scored items. \emph{Psychometrika}, 35(2), 179--197.
doi:10.1007/BF02291262

Bock, R. D., \& Mislevy, R. J. (1982). Adaptive EAP estimation of
ability in a microcomputer environment. \emph{Applied Psychological
Measurement}, 6(4), 431--444. doi:10.1177/014662168200600405

Bock, R. D., Thissen, D., \& Zimowski, M. F. (1997). IRT estimation of
domain scores. \emph{Journal of Educational Measurement}, 34(3),
197--211. doi:10.1111/j.1745-3984.1997.tb00515.x

Brinkhuis, M. J. S., \& Maris, G. (2019). Tracking ability: Defining
trackers for measuring educational progress. In B. P. Veldkamp \& C.
Sluijter (Eds.), \emph{Theoretical and practical advances in
computer-based educational measurement} (pp.~161--173). Cham: Springer.
doi:10.1007/978-3-030-18480-3\_8

Chang, H. (2004). \emph{Inventing temperature: Measurement and
scientific progress}. New York: Oxford University Press.

Choi, J. (2022). Cognitive Trait Model: Measurement model for mastery
level and progression of learning. \emph{Mathematics}, 10(15), 2651.
doi:10.3390/math10152651

Choi, J. (2026a). \emph{cogtraitmodel: Bounded-trait psychometrics with
the Anchored Logistic Family} (Version 0.2.2) {[}Computer software{]}.
doi:10.5281/zenodo.22031040

Choi, J. (2026b). \emph{Mastery scoring: A browser implementation of the
Cognitive Trait Model} (Version 1.0.0) {[}Computer software{]}.
doi:10.5281/zenodo.22050655

Choppin, B. H. L. (1985). Lessons for psychometrics from thermometry.
\emph{Evaluation in Education}, 9(1), 9--12.
doi:10.1016/0191-765X(85)90004-4

Corbett, A. T., \& Anderson, J. R. (1995). Knowledge tracing: Modeling
the acquisition of procedural knowledge. \emph{User Modeling and
User-Adapted Interaction}, 4(4), 253--278. doi:10.1007/BF01099821

Dempster, A. P., Laird, N. M., \& Rubin, D. B. (1977). Maximum
likelihood from incomplete data via the EM algorithm. \emph{Journal of
the Royal Statistical Society, Series B}, 39(1), 1--38.
doi:10.1111/j.2517-6161.1977.tb01600.x

Embretson, S. E. (1991). A multidimensional latent trait model for
measuring learning and change. \emph{Psychometrika}, 56(3), 495--515.
doi:10.1007/BF02294487

Gelman, A., Roberts, G. O., \& Gilks, W. R. (1996). Efficient Metropolis
jumping rules. In J. M. Bernardo, J. O. Berger, A. P. Dawid, \& A. F. M.
Smith (Eds.), \emph{Bayesian Statistics 5} (pp.~599--608). Oxford:
Oxford University Press.

Hastings, W. K. (1970). Monte Carlo sampling methods using Markov chains
and their applications. \emph{Biometrika}, 57(1), 97--109.
doi:10.1093/biomet/57.1.97

Hoffman, M. D., \& Gelman, A. (2014). The No-U-Turn Sampler: Adaptively
setting path lengths in Hamiltonian Monte Carlo. \emph{Journal of
Machine Learning Research}, 15(47), 1593--1623.

Ifrah, G. (2000). \emph{The universal history of numbers: From
prehistory to the invention of the computer}. New York: Wiley.

Kaplan, R. (2000). \emph{The nothing that is: A natural history of
zero}. Oxford: Oxford University Press.

Lord, F. M. (1953). The relation of test score to the trait underlying
the test. \emph{Educational and Psychological Measurement}, 13(4),
517--549. doi:10.1177/001316445301300401

Lord, F. M. (1975). The `ability' scale in item characteristic curve
theory. \emph{Psychometrika}, 40(2), 205--217. doi:10.1007/BF02291567

Lord, F. M. (1980). \emph{Applications of item response theory to
practical testing problems}. Hillsdale, NJ: Lawrence Erlbaum.

Lord, F. M., \& Novick, M. R. (1968). \emph{Statistical theories of
mental test scores}. Reading, MA: Addison-Wesley.

Lunn, D. J., Thomas, A., Best, N., \& Spiegelhalter, D. (2000). WinBUGS
--- a Bayesian modelling framework: Concepts, structure, and
extensibility. \emph{Statistics and Computing}, 10(4), 325--337.
doi:10.1023/A:1008929526011

Metropolis, N., Rosenbluth, A. W., Rosenbluth, M. N., Teller, A. H., \&
Teller, E. (1953). Equation of state calculations by fast computing
machines. \emph{Journal of Chemical Physics}, 21(6), 1087--1092.
doi:10.1063/1.1699114

Mislevy, R. J. (1986). Bayes modal estimation in item response models.
\emph{Psychometrika}, 51(2), 177--195. doi:10.1007/BF02293979

Noel, Y., \& Dauvier, B. (2007). A beta item response model for
continuous bounded responses. \emph{Applied Psychological Measurement},
31(1), 47--73. doi:10.1177/0146621605287691

Owen, R. J. (1975). A Bayesian sequential procedure for quantal response
in the context of adaptive mental testing. \emph{Journal of the American
Statistical Association}, 70(350), 351--356.
doi:10.1080/01621459.1975.10479871

Phan, D., Pradhan, N., \& Jankowiak, M. (2019). Composable effects for
flexible and accelerated probabilistic programming in NumPyro.
\emph{arXiv:1912.11554}.

Rasch, G. (1960). \emph{Probabilistic models for some intelligence and
attainment tests}. Copenhagen: Danmarks Paedagogiske Institut. (Expanded
ed., 1980, Chicago: University of Chicago Press.)

Robbins, H., \& Monro, S. (1951). A stochastic approximation method.
\emph{Annals of Mathematical Statistics}, 22(3), 400--407.
doi:10.1214/aoms/1177729586

Samejima, F. (1973). Homogeneous case of the continuous response model.
\emph{Psychometrika}, 38(2), 203--219. doi:10.1007/BF02291114

Spearman, C. (1904). ``General intelligence,'' objectively determined
and measured. \emph{American Journal of Psychology}, 15(2), 201--292.
doi:10.2307/1412107

van der Linden, W. J. (1979). Binomial test models and item difficulty.
\emph{Applied Psychological Measurement}, 3(3), 401--411.
doi:10.1177/014662167900300311

van der Linden, W. J. (1998). Bayesian item selection criteria for
adaptive testing. \emph{Psychometrika}, 63(2), 201--216.
doi:10.1007/BF02294775

von Davier, M., \& Lee, Y.-S. (Eds.) (2019). \emph{Handbook of
diagnostic classification models}. Cham: Springer.
doi:10.1007/978-3-030-05584-4

Wang, X., Berger, J. O., \& Burdick, D. S. (2013). Bayesian analysis of
dynamic item response models in educational testing. \emph{Annals of
Applied Statistics}, 7(1), 126--153. doi:10.1214/12-AOAS608

West, M., \& Harrison, J. (1997). \emph{Bayesian forecasting and dynamic
models} (2nd ed.). New York: Springer.

Wilcox, R. R. (1981). A review of the beta-binomial model and its
extensions. \emph{Journal of Educational Statistics}, 6(1), 3--32.
doi:10.3102/10769986006001003

\end{document}